\documentclass{aa}  

\usepackage{graphicx}
\usepackage{txfonts}

\usepackage[colorlinks=True]{hyperref}

\hypersetup{
    linkcolor=blue,  
    citecolor=blue,
    }

\newcommand{\kinemetry}{{\it Kinemetry}}

\newcommand{\msun}{\hbox{$\rm ~M_{\odot}$}}

\newcommand{\mbh}{{M$_{\bullet}$}}
\newcommand{\dg}{^{\circ}}
\newcommand{\arcdeg}{$^{\circ}$}

\newcommand{\kms}{km~s$^{-1}$}

\newcommand{\etal}{et al.~\/}

\newcommand{\fulloi}{[O\,{\small~I}]~$\lambda$6300}

\newcommand{\ha}{{H$\alpha$}}

\newcommand{\nii}{[N\,{\small~II}]}
\newcommand{\niione}{[N\,{\small~II}]~$\lambda$6548}
\newcommand{\niitwo}{[N\,{\small~II}]~$\lambda$6583}
\newcommand{\oi}{[O\,{\small~I}]}
\newcommand{\oione}{[O\,{\small~I}]~$\lambda$6300}
\newcommand{\oitwo}{[O\,{\small~I}]~$\lambda$6364}

\newcommand{\oiiione}{[O\,{\small~III}]~$\lambda$4959}
\newcommand{\oiiitwo}{[O\,{\small~III}]~$\lambda$5007}

\newcommand{\hanii}{H$\alpha$+[N\,{\small~II}]}
\newcommand{\fullhanii}{H$\alpha$+[N\,{\small~II}]~$\lambda\lambda$6548,6583}

\newcommand{\mgi}{Mg\,~$\lambda$5177}
\newcommand{\nai}{Na\,~$\lambda$5896}

\newcommand{\gal}{M87}

\newcommand{\Wbhm}{3.5 $\times\ 10^{9}$ \msun}

\newcommand{\python}{\textit{Python}}
\newcommand{\ppxf}{\textit{pPXF}}
\newcommand{\gist}{\textit{GIST Pipeline}}
\newcommand{\kinms}{\textit{KinMSpy}}

\newcommand{\sch}{Schwarzschild}
\newcommand{\nfm}{NFM}
\newcommand{\wfm}{WFM}

\newcommand{\iaa}[2]{#1 -- #2 $\AA$}
\newcommand{\ikms}[2]{#1 -- #2 \kms}
\newcommand{\idg}[2]{#1$\dg$--#2$\dg$}
\newcommand{\ias}[2]{#1$\arcsec$--#2$\arcsec$}
\newcommand{\ifas}[2]{0\farcs#1--0\farcs#2}
\newcommand{\fov}[2]{#1$\arcsec\times$#2$\arcsec$}

\begin{document} 

    \title{Revisiting the black hole mass of M87* using VLT/MUSE Adaptive Optics Integral Field Unit data I.}
    
    \subtitle{Ionized gas kinematics}

    \author{J. Osorno\inst{1}                           \thanks{E-mail: juanosorno@udec.cl},
        N. Nagar\inst{1}\thanks{E-mail: nagar@astro-udec.cl},
        T. Richtler\inst{1}, 
        P. Humire\inst{2}, 
        K. Gebhardt\inst{3}, 
        \and K. Gultekin\inst{4}
        }

    \institute{Departamento de Astronomía, Universidad      de Concepción, Barrio Universitario S/N,            Concepción, Chile
    \and
        Max-Planck-Institut f\"ur Radioastronomie, Auf dem H\"ugel 69, 53121 Bonn, Germany
    \and Department of Astronomy, University of Texas at Austin, Austin, TX 78712-1205, USA
    \and Department of Astronomy, University of 
        Michigan, Ann Arbor, MI 48109, USA
        }

   \date{Received March 31, 2023; accepted XXX XX, 2023}

 
  \abstract
   {Stellar dynamic-based black hole mass measurements of \gal\ are twice that determined via ionized gas kinematics; the former is closer to the mass estimated from the diameter of the gravitationally-lensed ring around the black hole.}
   {Using a deeper and more comprehensive ionized gas kinematic dataset, we aim to better constrain the complex morphology and kinematics of the nuclear ionized gas, thus gaining insights into the reasons behind the disagreement of the mass measurements.}
   {We use new Narrow Field Mode with adaptive optics, and Wide Field Mode integral field spectroscopic data from the \textit{Multi Unit Spectroscopic Explorer} instrument on the \textit{Very Large Telescope}, to model the morphology and kinematics of multiple ionized gas emission lines (primarily \fullhanii\ and \fulloi) in the nucleus of M87. 
   We use \kinemetry\ to fit the position angle, inclination, and velocities of the sub-arcsec ionized gas disk.
    We use \kinms\ to create simulated datacubes across a range of black hole masses and disk inclinations, and parameterized the differences
    of the resulting residual (observed minus simulated) velocity maps, in order to obtain the best fit model.}
   {The new deep dataset reveals complexities in the nuclear ionized gas kinematics which was not seen in earlier sparse and shallower \textit{Hubble Space Telescope} spectroscopy. Several ionized gas filaments, some with large flow velocities, can be traced down into the projected sphere of influence. However, not all of these truly pass close to the black hole.
   Additionally, we find evidence of a partially-filled biconical outflow, aligned with the jet, with radial velocities up to 400 \kms. 
   The sub-arcsec rotating ionized gas 'disk' is well resolved in our datacubes. 
   The velocity isophotes of this disk are twisted and the position angle of the innermost ($\lesssim$5 pc) gas disk tends toward a value perpendicular to the 
    radio jet axis.
   The complexity of the nuclear morphology and kinematics (the mix of a warped disk with spiral arms, large linewidths, strong outflows, and filaments crossing the black hole in projection)
   precludes the measurement of an accurate black hole mass from the ionized gas kinematics. 
    Two results (each relatively weak; but together more convincing) support a high mass ($\sim$6.0 $\times$ 10$^{9}$ \msun) black hole in a low (i $\sim$25$\dg$) inclination disk rather than a low mass ($\sim$3.5 $\times\ 10^{9}$ \msun) black hole in a i = 42$\dg$ disk: 
   (a) \kinemetry\ fits to the sub-arcsec disk support inclinations of $\sim$20$\dg$--25$\dg$, rather than 42$\dg$;
   (b) velocity residual  (observed - simulated) maps with slightly smaller residuals are found for the former case. 
   The specific (sub-Keplerian) \textit{radiatively inefficient accretion flow} (RIAF) model earlier proposed to reconcile the mass measurement discrepancy was also tested: the sub-Keplerian factor used in this model is not sufficiently small to make a high mass black hole in a RIAF inflow masquerade as a low mass black hole in a Keplerian inflow. In general Keplerian disk models perform significantly better than the RIAF model when fitting the sub-arcsec ionized gas disk.}
   {A disk inclination close to 25\arcdeg\ (rather than the previously posited 42\arcdeg) for the nuclear gas disk, and the warp in the sub-arcsec ionized gas disk, help to reconcile the contradictory nature of key earlier results: (a) the mass discrepancy between stellar and ionized gas black hole masses (our results support the former); (b) the mis-orientation between the axes of the ionized gas disk and the jet (we find these aligned in both 2- and 3-dimensions). Further, we identify a previously unknown 400 \kms\ (partially-filled) biconical outflow along the (3-dimensional) jet axis, and show that the velocities of the two largest ionized gas filaments at 8\arcsec--30\arcsec\ nuclear distances can be explained primarily by rotation in the extension of the (inclination $\sim$25\arcdeg) nuclear ionized gas disk.
   }

   \titlerunning{Ionized gas kinematics around M87*}
   \authorrunning{Osorno, et al.}
   \maketitle
%

\section{Introduction}
\label{intro}

There is increasing evidence that all the galaxies with a bulge component have a nuclear supermassive black hole \citep[SMBH; e.g., ][]{sag16}. Direct measurements of SMBH masses - available for only $\sim$200--250 galaxies - are enabled via molecular gas \citep[e.g.,][]{bar16}, ionized gas  \citep[e.g.,][]{mul11} and water vapor maser \citep[e.g.,][]{gao17} kinematics, stellar dynamics \citep[e.g.,][]{rus11},
and reverberation mapping \citep[RM; e.g.,][]{ben10}. 
For a review of these methods and implications for black hole and galaxy co-evolution see e.g, \citet{kor13}.

For this relatively small and biased sample of SMBH measurements, the SMBH mass (\mbh) appears to be related to several properties of the host galaxy, such as the bulge luminosity \citep[][]{kor01,mar03,gul09} and the bulge stellar velocity dispersion \citep[][]{fer00,gul09,sag16}. Additionally, the so-called 'single-epoch RM' method (using the width of a broad emission line and the luminosity of the nearby continuum in broad line active galactic nuclei or AGN) can be used to estimate the SMBH mass. These scaling relationships have been used to understand the role of the black holes in the formation and growth of their host galaxies \citep[e.g.,][]{dim05}. 

Measurements of a SMBH mass using different techniques can result in inconsistent results \citep[e.g.,][]{wal13,wisdomvii}.
A famous example is NGC 4486 (\gal) where the measurement from stellar dynamics \citep[][hereafter G11]{geb11} is twice the value from ionized gas kinematics \citep[][hereafter W13]{wal13}, with the two values differing by more than the 3$\sigma$ of each quoted measurement. 

NGC4486 (\gal) is a giant elliptical galaxy located in the center of the Virgo cluster. It has a prominent relativistic jet which has been well studied on scales of 10 gravitational radii \citep[][]{ehtc19} out to $\sim$40 kpc from the nucleus \citep[e.g., ][]{Owen2000}.
The jet is projected on the sky with a position angle (PA) of 288\arcdeg\ \citep[][]{wal18}. The orientation and inclination of the nuclear ionized disk was constrained to PA = 45\arcdeg\ and inclination (i) $\sim$42\arcdeg\ by \citet{wal13}. Given this, \citet{jet19} argued that the jet has an inclination of 18\arcdeg\ to the line of sight and found that the jet and gas disk axes are misaligned by at least 11\arcdeg\ (with a typical misalignment of 27\arcdeg). 

There are several reasons to use \gal\ as a test case for black hole mass measurement (in)consistency from different methods: it is one of the brightest galaxies in the sky; as a massive cD elliptical with a large velocity dispersion, it is expected to host a massive $\geq$ $10^{9}$\msun\ SMBH; and it hosts one of the most studied and well-known 'radio-loud' AGN. 

The first attempts to measure \mbh\ in \gal\ were made by \citet{sar78} and \citet{you78}, who suggested a black hole mass of about $10^{9}$\msun. In companion papers, \citet{for94} presented \fullhanii\ emission-line images of the M87 nucleus using data from the WFPC2 camera aboard the \textit{Hubble Space Telescope}  (HST), and \citet{har94} presented spectra at six positions in the M87 nucleus using data from the Faint Object Spectrograph aboard HST. While the nuclear emission line region was  clearly not an isolated ionized disk, their isophotal fits to the emission line flux distribution in the nuclear arcsecond, the presence of trailing spiral arms, and the large emission line velocity gradients between apertures, led them to posit a nuclear ionized gas disk in PA $\sim$1\arcdeg--9\arcdeg, inclination $\sim$42\arcdeg, and a black hole mass of 2.4 $\times$ 10$^9$ \msun\ assuming a distance of 15 Mpc to M87. \citet{mac97} used three long slit spectra from the Faint Object Camera aboard HST and found that the kinematics are best explained by a 3.2 $\times$10$^9$ \msun\ black hole (for a distance of 15 Mpc and i = 51\arcdeg), but noted that the disk inclination is the most uncertain parameter, with likely values between 47\arcdeg and 65\arcdeg. 

G11 used \sch\ modeling of  observed stellar brightness and line of sight velocities to measure a  black hole mass of (6.6 $\pm$ 0.4) $\times\ 10^{9}$ \msun\ for a distance of 17.9 Mpc. They used data from the Integral Field Spectrograph (NIFS) on the Gemini North Telescope with adaptive optics (AO) correction, combined with extensive  kinematics out to large radii. The large scale stellar kinematics came from two sources: SAURON integral field unit (IFU) data from \citet{ems04} and VIRUS-P IFU data from \citet{mur11}. 

\citet{wal13} measured the black hole mass of \gal\ via the kinematics of the \fullhanii\ emission lines using data from the Space Telescope Imaging Spectrograph (STIS) aboard HST. Their spectral data were taken in five parallel slits in PA = 51\arcdeg, separated by 0\farcs1 in the plane of the sky, and with the central slit crossing the galaxy nucleus. They interpreted the ionized gas emission lines as coming from a rotating disk, with PA = 45\arcdeg\ and i = 42\arcdeg. With these parameters, and eliminating data from pixels very close to the nucleus where the \ha\ and \nii\ lines are highly blended, they derived a \mbh\ of 3.5$^{+0.9}_{-0.7}$ $\times\ 10^{9}$ \msun\ for a galaxy distance of 17.9 Mpc.

Recently, the black hole mass of \gal\ was constrained using data from the 2017 campaign of the \citet[EHT Collaboration,][]{ehtc19}. A SMBH enveloped in radio emitting plasma is expected to be have a bright gravitationally lensed ring of emission surrounding the black hole 'shadow'. The diameter of ring is predicted to be $\sim$9.6--10.4 G\mbh/$c^{2}$, relatively independent of the SMBH spin \citep{Johannsen2010,Gralla2020}. The ring diameter measured by the EHT, together with standard General Relativity, imply a \mbh\ of 6.5 $\pm$ 0.2 (statistic) $\pm$ 0.7 (systematic) $\times\ 10^{9}$ \msun\ for a distance to the galaxy of 16.8$^{+0.75}_{-0.66}$ Mpc. Scaled to this distance, the G11 and W13 SMBH masses are, respectively, 6.14$^{+1.07}_{-0.62}$ $\times\ 10^{9}$ \msun\ and 3.45$^{+0.85}_{-0.26}$ $\times\ 10^{9}$ \msun\ \citep{ehtc19}. The EHT result is thus compatible with G11 but not with W13.
More recently, the EHT Collaboration used the size of the ring around the SMBH in the Galaxy (Sgr A$^*$), together with standard General Relativity, and found an excellent agreement between the 'BH shadow' derived SMBH mass with the previous high precision (resolved) stellar dynamics mass \citep{ehtc21}. This latter result lends additional confidence to the EHTC derived SMBH mass of M87.

Most recently, \citet{lie23} studied the stellar kinematics of \gal\ with data from the integral field spectroscopic at the Keck II Telescope, concluding that the galaxy is strongly triaxial (there is a misalignment of 40$\dg$ between the kinematic axis and the photometric major axis from $\sim$5 kpc and outward). They derived a SMBH mass of 5.37$^{+0.37}_{-0.25}\times 10^{9}$\msun\ (for a galaxy distance of 16.8 Mpc), using \sch\ orbit modeling with a varying mass-to-light ratio (M/L). 

To summarize previous SMBH mass measurements (and assuming a distance of 16.8 Mpc to M87): stellar dynamics and EHTC measurements suggest a more massive SMBH (5.4--6.5 $\times\ 10^{9}$ \msun) while gas kinematics suggest a significantly lower SMBH mass: 3.5 $\times\ 10^{9}$ \msun. Two immediate explanations for the discrepancy of the ionized gas kinematic value are a lower inclination for the nuclear ionized disk, and sub-Keplerian rotation of ionized gas \citep{jet19}; both are explored in this work. While we explore a full parameter space of SMBH mass 2--8 $\times\ 10^{9}$ \msun\ and inclination 20--50\arcdeg, some pv diagrams and residual velocity maps use an illustrative 'high mass black hole' (hereafter HBH; 6.0 $\times\ 10^{9}$ \msun), a rough mean value of the three high mass SMBH measurements) and a 'low mass black hole' (hereafter LBH; 3.5 $\times\ 10^{9}$ \msun). Further, we also use illustrative models with disk inclination i = 42$\dg$ (the value derived in \citet{wal13}) and i = 25$\dg$. For example,  the HBH i25 model is a HBH black hole in a disk with inclination 25$\dg$.

A more comprehensive understanding of the complexities of the ionized gas kinematics requires integral field unit (IFU) spectroscopy. Indeed, the morphology of the ionized gas in \gal, on arcseconds to tens of arcseconds scales, is filamentary and complex, with evidence of multiple velocity components. The ionized gas maps of \citet[][]{bos19} show  filaments extending from the nucleus up to $\sim$3 kpc to the NW and $\sim$8 kpc to the SE. They used data from the IFU Multi Unit Spectroscopic Explorer \citep[MUSE,][]{muse} on the \textit{Very Large Telescope} (VLT), in its Wide Field Mode (\wfm), to model the ionized gas kinematics in the inner arcmin, and showed that the filaments have perturbed kinematics with velocity differences  of $\sim$\ikms{700}{800}, but a fairly uniform velocity dispersion of $\sim$100 \kms\ over the filaments. 

To better understand the difference between the M87 SMBH masses measured using ionized gas, stellar dynamics, and EHT ring-size, we have obtained new observations of \gal\ with MUSE in its AO Narrow Field Mode (\nfm) mode. The data have significantly higher spatial resolution than the WFM data used by \citet{bos19} and both a higher spatial resolution and better signal-to-noise ratio (SNR) than all previous HST/STIS data. Further, unlike the data from STIS, the MUSE \nfm\ cube fully samples the nuclear \fov{8}{8} region and provides a larger wavelength coverage (thus additional emission and absorption lines), and can be used to model both the stellar dynamics and ionized gas kinematics. Additionally, previous MUSE \wfm\ observations of \gal\ over the central 1\arcmin, allow constraints on the larger scale morphology and kinematics/dynamics.

In this work, we use \wfm\ and \nfm\ data to constrain the nuclear ionized gas morphology and kinematics in order to better understand the mismatch between ionized gas and stellar dynamic black hole mass measurements. In a forthcoming work (Osorno et al., in prep.), we will revisit the stellar dynamics black hole mass measurement using the stellar velocity line of sight velocity distributions measured in the same datacubes. The dust and ionized gas structures, and their relationship with the prominent jet, will be explored in Richtler et al. (in prep.).

For clarity, we mention the a-priori parameters used in our analysis here.
As mentioned above, the EHT Collaboration adopted a distance to \gal\ of 16.8 Mpc (thus a scale of 0.081 kpc per arcsec), and we use this distance in our work.
G11 and W13 used a distance of 17.9 Mpc to derive their SMBH masses; these masses were scaled to the EHT distance in \citet{ehtc19}, and we use these updated values (listed above).
The heliocentric recessional velocity adopted by NED for \gal\ is 1284 \kms\ (from stars). The radial velocity listed in SIMBAD is 1256 \kms. The best fit of W13 to the ionized gas kinematics in HST/STIS data yielded a radial velocity of 1335 \kms. 
Our fit to the stellar absorption lines in an annular radius between 2\arcsec\ and 24\arcsec\ in the MUSE \wfm\ cube yields a radial velocity of 1313 \kms. Our equivalent fit to the full field of view (FOV) of the \nfm cube yields a radial velocity of 1301 \kms. The rotational component of the diffuse ionized gas (i.e., not considering the velocity of the spiral arms) in the inner 1\arcsec\ gas disk in our \nfm\ cube is best centered using a radial velocity of 1260 \kms. In this work, unless explicitly mentioned otherwise, we use 'zero' radial velocities of 1313 \kms\ and 1260 \kms\ for the stellar and ionized gas kinematics, respectively. 

The paper is distributed in the following sections: observations and data processing in Sect. \ref{obsanddata}; 
our programs and modeling procedure in Sect. \ref{metandsoft}; 
relevant stellar kinematics and properties in Sect. \ref{stkin}; 
the ionized gas moment maps in Sect. \ref{secionmom};
the nuclear ionized disk geometry in Sect. \ref{igkinemetry}; 
ionized gas filaments in Sect. \ref{igfilaments}; 
ionized gas pv diagrams along major and minor axes in Sect. \ref{igpvs};
residual velocity maps in Sect. \ref{rvmaps};
constraints on black hole mass in Sect. \ref{igkinms};
the biconical outflow in Sect. \ref{igoutflow};
and finally the discussion and conclusions of this study in Sect. \ref{discussion}.

\section{Observations and Data Processing}
\label{obsanddata}

Observations were made with the MUSE IFU mounted on the VLT \citep{muse}. MUSE covers a wavelength range of \iaa{4650}{9350} with a wavelength sampling of 1.25 $\AA$ per pixel in both \wfm\ and \nfm\ modes. The spectral resolution is R = 3026 at 7000\AA. The \wfm\ mode has a FOV of \fov{60}{60}, and a spatial sampling of 0\farcs2 per pixel.
The \nfm\ mode has a FOV of \fov{8}{8} with a spatial sampling of 0\farcs0252 per pixel.
Since \nfm\ data are obtained in adaptive optics mode with Sodium laser artificial guide stars, no data are obtained in the wavelength range of \iaa{5780}{6050}.  

The \wfm\ mode observations were made on June 28, 2014, during a MUSE Science Verification run. Data were taken at airmass 1.3, with a DIMM seeing of 0\farcs9.
The \nfm\ mode observations were made on February 2, 2020, as part of project ID 0103.B-0581 (P.I. Nagar). Data were taken at airmasses between 1.25 and 1.5, with DIMM seeing values of 0\farcs4 to 0\farcs5. 

The \wfm\ data were processed with the standard MUSE pipeline (v1.6.1) and the calibrated datacubes were hosted on the ESO Archive (Processed Data Science Portal). A single processed datacube, with a total on-sky exposure time of 1800 seconds, was downloaded from this archive. 
The \nfm\ data were processed by the standard MUSE pipeline  (v2.8.4) and the calibrated datacubes were hosted on the same ESO Science Portal. Three processed datacubes, each with 2100 seconds of on-sky exposure, were downloaded from this archive. A fourth observation, which was aborted after 83 seconds of exposure, was not used. 
The three NFM datacubes were aligned using the position of the bright nucleus and jet knots (one of the cubes had to be shifted by 2 pixels to the E and 1 pixel to the N) and combined into a single 6300 second exposure datacube. 
The \nfm-AO mode point-spread-function (PSF) is best defined as a double Moffat function: one for the core and the other for the wings.
Given that M87 is relatively northern for the VLT and has to be observed at relatively large airmass, the FWHM of the PSF core is expected to be better than 100 mas
(see the MUSE \textit{Exposure Time Calculator}\footnote{\url{https://www.eso.org/observing/etc/bin/gen/form?INS.NAME=MUSE+INS.MODE=swspectr}}
and MUSE User Manual\footnote{\url{https://www.eso.org/sci/facilities/paranal/instruments/muse/doc.html}}). 
With no stars in the FOV, we were unable to measure the true PSF in our dataset. However, three globular clusters in M87
can be distinguished in a continuum image of the \nfm\ cube after subtracting a model galaxy bulge. The FWHM of the profiles of these globular clusters range between 82 and 93 mas.

The final \nfm\ datacube provides full areal coverage of the inner \fov{8}{8}, at a spatial resolution better than HST, a depth significantly better than the previous observations of W13, and an almost complete coverage of the optical wavelength range. Appendix \ref{app:stis-muse} compares the two datasets, with  Fig. \ref{fig:stismuse} highlighting the large improvement the new dataset brings. 
 
\section{Methods and software}
\label{metandsoft}

In this section we very briefly describe the process and programs used to obtain the stellar and ionized gas kinematics, and the models used in our analysis. A more detailed description of these can be found in Appendix \ref{app:methods}. 

Most codes used, both publicly available and our own, are in \python, and we extensively use the 'astropy' library.
We used the version 3.1.0. of the \gist\ software \citep{gist} which integrates programs for deriving the stellar and ionized gas kinematics of galaxies from optical data cubes. 
Stellar kinematics, and stellar population properties, were constrained using the version of the penalized Pixel Fitting Technique \citep[\ppxf;][]{ppxf1,ppxf2} included in the \gist. Different binning techniques were used to bin spaxels in order to increase the SNR of the spectra, and spaxels which include the jet were not included in the analysis. For details, see Appendix \ref{app:methods}.

Ionized gas kinematics was derived in two ways. First, using \textit{GANDALF} \citep{gandalf1,gandalf2,gist} within the \gist. Here, in each spectrum, a simultaneous fit to both absorption (from a previous \ppxf\ run) and emission lines, is used to obtain an 'emission-line-only' spectrum. Since  spatial binning was required to increase the reliability of the absorption line fits, the resulting emission line kinematics are obtained in the same bins. While these GIST results were used to check the results below, we do not show them in the figures of this work.
Second, to maintain the full intrinsic resolution of the datacube, for each emission line of interest we fit a single Gaussian (for the emission line), plus a first order polynomial (for the continuum), to obtain the intensity, velocity, and velocity dispersion of the emission line in each spaxel. In the case of emission line complexes with wavelengths close enough to be blended in the datacube, e.g., \fullhanii, we simultaneously fit each emission line with a single Gaussian. Velocities and dispersions of the different emission lines were tied, and some line ratios were fixed, in these multiple Gaussian fits (see Appendix \ref{app:methods}).

We also build position-velocity (pv) diagrams to reveal the full detail of the ionized gas kinematics, especially in the presence of multiple components. These are extracted from the data cubes along both straight slits and curved pseudo-slits.  

In Sects. \ref{igpvs}, \ref{rvmaps}, and \ref{igkinms}, we compare observed velocity maps and pv diagrams directly to analytic kinematic models and to simulated cubes based on analytic kinematic models. In the 'Keplerian' model, we assume that the ionized gas is rotating in a thin disk, with a given PA and inclination, under the influence of both the potential of the nuclear black hole and the galaxy. The rotational velocity of the gas ($v_{gas}$) is
\begin{equation}
    \label{gasvel}
    v_{gas}(r)=\sqrt{\frac{G\left[M_{\bullet}+M_{bulge}(r)\right]}{r}},
\end{equation}
where $r$ is the nuclear distance, $G$ the gravitational constant, $M_{\bullet}$ the black hole mass and $M_{bulge}$ is the total mass of the stars within a sphere of radius $r$ centered on the nucleus. For its part, $M_{bulge}$ is derived with the following equation
\begin{equation}
    \label{mbulge}
    M_{bulge}(r)=\int_{0}^{r}4\pi r'^{2}\Upsilon(r')\nu(r')dr',
\end{equation}
where $\Upsilon$ is the radially varying mass to light ratio (M/L) of the stars, $\nu$ is the volume luminosity density of the stars, and $r'$ is the integration variable. The V-band luminosity density of \gal, as a function of radius, is taken from Fig. 1 of \citet{geb09}. The values in this plot were extracted with version 3.5.1 of the \textit{Plot Digitizer Online Application}\footnote{\url{https://plotdigitizer.com}}. As the luminosity density profile is not an analytic function, the integral in Eq. \ref{mbulge} was performed numerically. Given the large black hole mass, the galaxy potential is significant only at r $\gtrsim$ 2\arcsec. The radially varying M/L ratio in the V-band is taken from \citet{lie23}, who fit the R-band M/L variation determined by \citet{sar18} with a logistic equation, obtaining e.g. values of V-band M/L of 8.65 and 3.46 in the central and the outer part of \gal, respectively. 

The line-of-sight (LOS) velocity of the ionized gas in a given spaxel or along a slit was obtained by projecting $v_{gas}$ with the following equation
\begin{equation}
    \label{velproj}
    v_{los}=v_{gas}\sin{i}\cos{\theta},
\end{equation}
where $i$ is the inclination of the disk with respect to the plane of the sky, and $\theta$ is the angle (in the plane of the disk) between the major axis of the disk and the deprojected position angle of the slit. 

We also compare our observables to the \textit{radiatively inefficient accretion flow} (RIAF) model which \citet{jet19} advanced as a potential explanation for the ionized gas derived black hole mass being lower than the stellar derived one in M87. This model considers the presence of sub-Keplerian rotational in the ionized gas, with support compensated with a radial outflow in the disk. 
\citet{jet19} parameterized the radial and circular velocities as, respectively, $v_{r}=-\alpha v_{kep}$ and $v_{\phi}=\Omega v_{kep}$, where $v_{kep}$ is the Keplerian velocity (Eq. \ref{gasvel}).
While $\alpha$ and $\Omega$ in RIAFs can have a range of values, we follow the specific RIAF parameters used by \citet{jet19}: $\alpha=\sqrt{0.1}$, $\Omega=\sqrt{0.7}$, and a velocity dispersion profile $\sigma=\sqrt{f}v_{kep}$, with $f=0.1$.

We use the program \kinemetry\footnote{\url{http://davor.krajnovic.org/software/}}\ \citep{kinemetry} to constrain the best fit PA and inclination of the ionized gas disk. \kinemetry\ fits the line of sight velocity in concentric ellipses, and provides the best fit PA, inclination, circular velocity, and non-circular velocity components for each fitted ellipse (i.e. at each radius). Velocity field fitting can be done with the PA and inclination as free or fixed parameters.

We create simulated emission line MUSE cubes using the \kinms\ software \citep{kinms}. While this was designed to simulate molecular gas cubes, since the third axis of the cube is in velocity space, it is easily adapted to MUSE datacubes. \kinms\ allows our analytical rotation models to be convolved with the same spatial and spectral broadening and sampling of the observed dataset.
When creating a simulated emission line cube for lines from the nuclear disk, we used the program \textit{skySampler} in \kinms\ to replicate the surface brightness of the emission lines; \textit{skySampler} distributes the $\sim$2 million simulated disk clouds in order to best fit the intensity image of the emission line. The velocity differences  between the observed and simulated velocity maps are compared using several metrics and masks in order to obtain the best fit simulated cube, thus analytic model.

The 'toy' nuclear ionized outflow in Sect. \ref{igoutflow} is also simulated with \kinms. This is modeled using one or two filled cones with vertices in the nucleus. We define our coordinate system with the $x$ axis positive to the E, the $y$ axis positive to the N, the observer along the positive z axis, and the origin at the nucleus. 
The cones inclination to the observer, their opening angles, lengths, and their radial outflow velocity, were adjusted - by eye - for a good general agreement with the observed velocity maps. 

\section{Stellar Properties from the \wfm\ datacube}
\label{stkin}

\begin{figure*}
\centering
\includegraphics[scale=0.5,clip,angle=0]{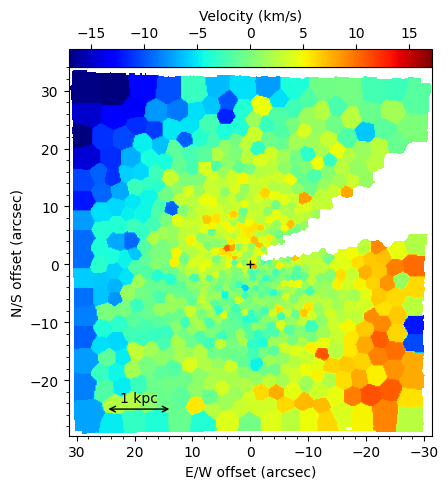} 
\includegraphics[scale=0.5,clip,angle=0]{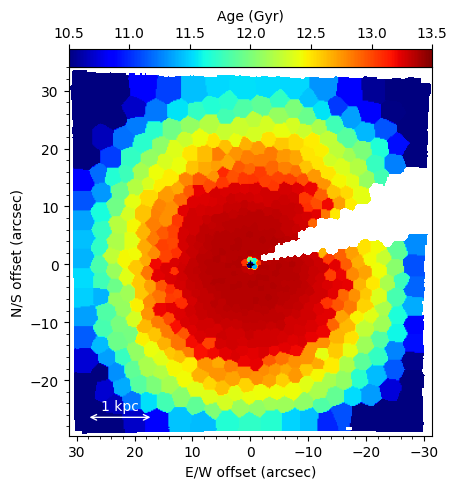}
\includegraphics[scale=0.5,clip,angle=0]{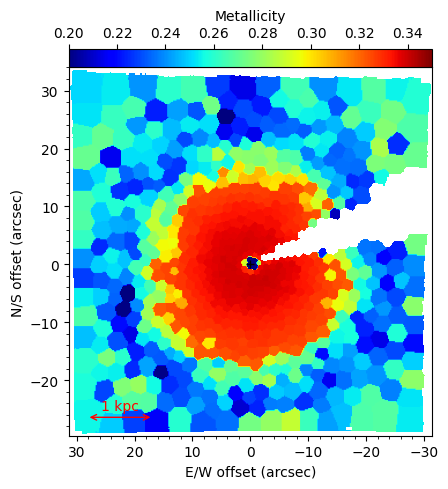} 
\caption{Maps of stellar properties derived with the \gist\ run on the MUSE \wfm\ datacube. 
Left: stellar velocity map; zero velocity corresponds to a radial velocity of 1313 \kms. 
Middle: map of the weighted mean stellar age of the best-fit templates.
Right: map of the weighted mean metallicity of the best-fit templates.
These maps were derived using \gist\ over a wavelength range of \iaa{5050}{6000}, with Voronoi binning used to obtain a SNR $\geq$ 300 in each bin. 
All spaxels with significant jet emission were masked in the datacube to avoid confusion.} 
\label{fig:stmoms}
\end{figure*}

A full analysis of the stellar dynamics and \sch\ modeling of the line of sight stellar velocity distribution to constrain the black hole mass, is deferred to a future work (Osorno \etal, in prep.). Here we only present three stellar-related maps most relevant to our interpretation of the ionized gas kinematics. 

Stellar moment maps and weighted mean ages and metallicities of the best-fit stellar templates were constructed with \ppxf\ within the \gist\ (see Appendix \ref{app:methods} for details). The left panel of Fig. \ref{fig:stmoms} shows the velocity map derived from the \wfm\ cube. A systemic (radial) velocity of 1313 \kms\ was used as 'zero' velocity.   
As seen in the \wfm\ velocity map, and as earlier noted by \citet{ems14}, there are two rotation patterns: one of $\pm$5 \kms\ in the inner $\sim$18\arcsec\ with redshift to the NE, and the other of $\pm$20 \kms\ (in the opposite direction, i.e. redshift to the SW) and going out to beyond the FOV of the cube. The velocity dispersion map (not shown) is symmetric with values $\sim$250 \kms\ in a ring of nuclear radius $\sim$20\arcsec, and steadily increasing to more than 400 \kms\ with decreasing nuclear distance. Given that, as expected, stellar dispersion dominates stellar rotation in the nucleus; therefore, we will ignore these $\pm$5 \kms or $\pm$20 \kms\ of stellar rotational velocities when modeling the gas kinematics in the next sections. 

The middle and right panels of Fig. \ref{fig:stmoms} show the maps of weighted stellar age and metallicity from the \wfm\ cube. While the fitted stellar population gets older when approaching the nucleus, the stellar age and metallicity do not vary significantly out to $\sim$12\arcsec, with values between 13--13.5 Gyr and 0.32--0.35 respectively (except in the innermost arcsec, where the large dispersion, and the nuclear jet emission, potentially confuse the fits). Outside $\sim$12\arcsec, the stellar age decreases rapidly with increasing nuclear distance, reaching values below 10.5 Gyr. The metallicity also decreases with radial distance between 12\arcsec\ and $\sim$23\arcsec\ (values between 0.2--0.24) and then slightly increases to values between 0.24--0.3. It is worth noting the steep change in the metallicity at radius $\sim$12\arcsec, roughly the outer radius of the counter-rotating core. 
For a V-band M/L ratio of 4 (as used by W13) and a G11 black hole mass, the enclosed galaxy mass equals the black hole mass at radius $r\sim$6\farcs75. Our use of a higher V-band M/L ratio in the nucleus, following \citet{lie23}, still does not affect our results in the inner arcsec of the ionized gas disk, and is mainly relevant in the analysis of the kinematics of ionized filaments at $\gtrsim$2\arcsec.

\section{Ionized gas: Moment maps}
\label{secionmom}

\begin{figure*}[h!]
\centering
\includegraphics[scale=0.52,clip,angle=0]{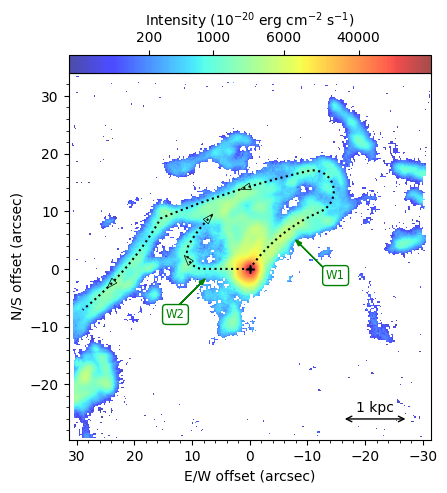}
\includegraphics[scale=0.52,clip,angle=0]{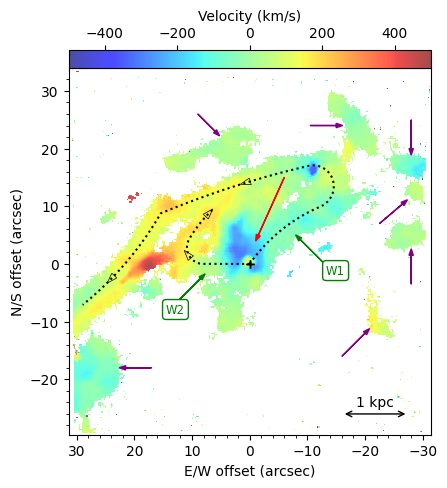}
\includegraphics[scale=0.52,clip,angle=0]{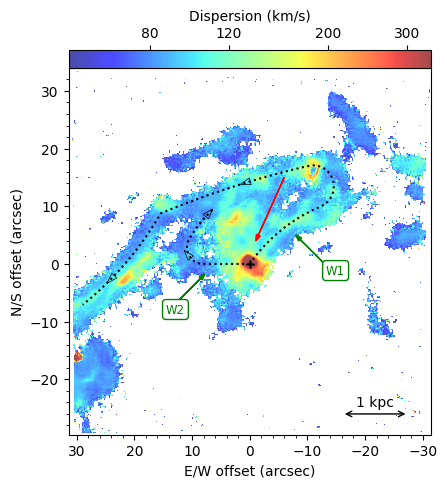}
\includegraphics[scale=0.52,clip,angle=0]{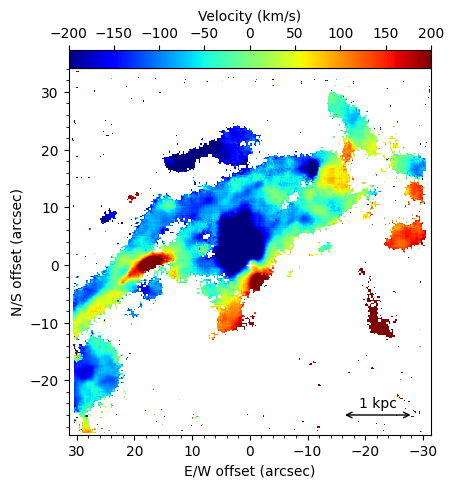}
\includegraphics[scale=0.52,clip,angle=0]{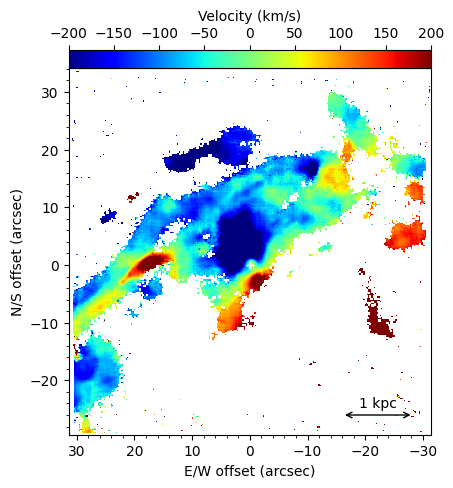}
\includegraphics[scale=0.52,clip,angle=0]{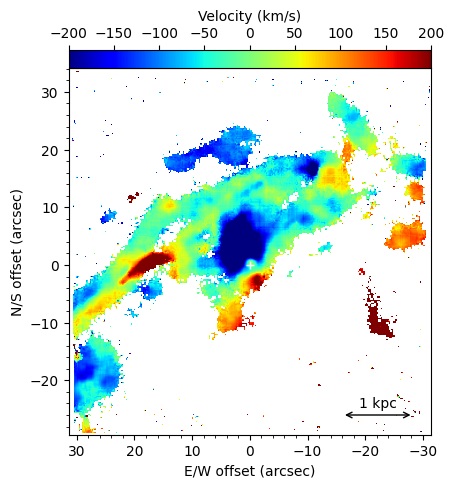}
\caption{Moment and residual maps for the \niitwo\ line from the MUSE \wfm\ datacube. 
Top: moment maps of the \niitwo\ line as derived from Gaussian fits to the datacube. Left to right are the total intensity, velocity, and dispersion maps. Two ionized gas filaments (black dotted lines) are indicated by green arrows and the labels W1 and W2; these dotted lines are the pseudo-slits along which the pv diagrams of Fig. \ref{fig:pvwfm} are extracted. Several ionized gas bubbles are indicated by purple arrows and the position of the outflow is indicated by a red arrow. 
Bottom: residual velocity maps of the \niitwo\ line in the datacube. These maps, discussed in Sect. \ref{wfmfil}, show the residual velocity field after subtracting a model of a rotating thin nuclear disk (in a SMBH plus stellar potential) for a black hole and nuclear disk with the following parameters (\mbh, disk inclination): left: 6.0 $\times\ 10^{9}$ \msun\ and 42$\dg$; middle: \Wbhm\ and 42$\dg$; and right: 6.0 $\times\ 10^{9}$ \msun\ and 25$\dg$ (our preferred model; see Sect. \ref{wfmfil}); all with disk position angle of 45$\dg$.} 
\label{fig:wfmmoms}
\end{figure*}

\begin{figure*}[h!]
\centering
\includegraphics[scale=0.55,clip,angle=0]{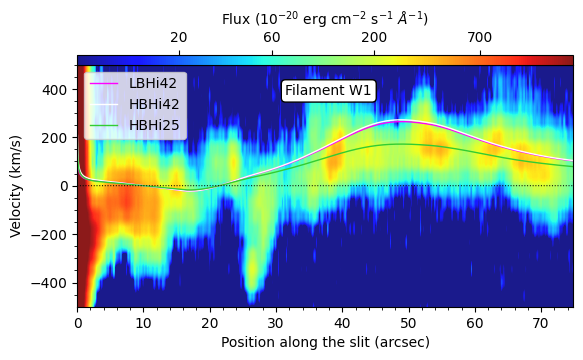} 
\includegraphics[scale=0.55,clip,angle=0]{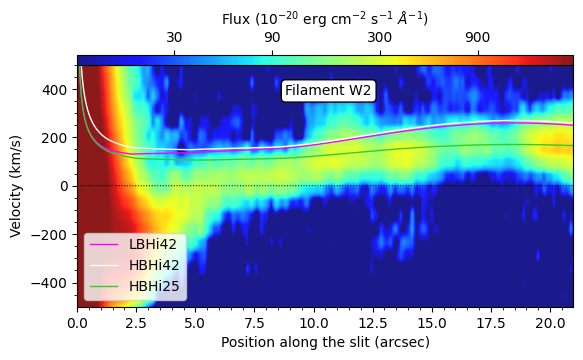} 
\caption{Position-velocity diagrams of the \niitwo\ line for pseudo-slits along the gas filaments W1 and W2 identified  in Fig. \ref{fig:wfmmoms}. The $x$ axis zero position of each filament corresponds to the black cross in Fig. \ref{fig:wfmmoms}. The three curves are the expected disk rotation velocities with the parameters described in Sect. \ref{wfmfil} and Fig. \ref{fig:wfmmoms}, with colors as listed in the insert. 
At offsets beyond 10\arcsec, the velocities of both filaments are best fit with the i = 25\arcdeg\ model. This figure is discussed in Sect. \ref{wfmfil}, but placed here for easy comparison with Fig. \ref{fig:wfmmoms}.
}
\label{fig:pvwfm}
\end{figure*}

The unprecedented depth, spatial resolution, and full areal coverage of the MUSE \wfm\ and \nfm\ cubes reveal new complexities in the nuclear ionized gas kinematics. 
Since the \niitwo\ emission line is the brightest in both the \nfm\ and \wfm\ cubes, we use this line extensively in our analysis. However, in the innermost $\sim$1\farcs5, the emission lines are very broad and have multiple velocity components, so we are unable to clearly separate the \ha\ and \nii\ lines in our Gaussian fits. We thus also rely on the \oione\ emission line in this area, since this is a relatively isolated emission line and is bright enough to derive reliable kinematics in the inner arcsec. 

\subsection{\ha\ + \nii\ \wfm\ Moment maps}
\label{secmomnfmha} 

Ionized gas maps of the central arcmin of \gal\ were previously presented by \citet{bos19} using the same \wfm\ data cube used here; for these maps we thus  only briefly review relevant known results, and discuss features relevant to the black hole mass determination. 
Figures \ref{fig:wfmmoms} (top row) show the intensity, velocity, and dispersion maps of \niitwo\ obtained from the \wfm\  data cubes. The velocity maps assume a systemic velocity of $1260$ \kms, and a sigma clipping was applied to remove bad pixels. 
Several features are worth noting in these moment maps. The two black dotted lines, indicated with green arrows, trace the two primary ionized gas filaments which enter the nucleus (in projection).
The filament labeled "W1" is the most prominent nuclear filament and extends beyond the MUSE \wfm\ FOV to the W \citep{gav00,wer10,bos19}. 
The filament marked "W2" also crosses the nucleus but from the E, and connects to filament W1 about 10\arcsec\ to the NE of the nucleus. There is an additional highly redshifted connection from the midpoint of filament W2 to filament W1. 
Additional shorter filaments and clouds, indicated in the velocity map with purple arrows, show velocities relatively different from that of filament W1. The velocity map shows a $\sim$6\arcsec\ diameter region to the N of the nucleus (indicated with the red arrow in the velocity and dispersion maps) that is $\sim$200 \kms\ blueshifted from systemic. We argue below that this is from radial outflow whose axis is aligned with the jet, which adds additional complexity to the nuclear kinematics. 
In the innermost arcsecond one can already distinguish the posited disk-like rotating component with redshifts (blueshifts) up to $\sim$400 \kms\ to the NW (SE). 

The dispersion map shows high dispersion to the NE of the nucleus; we argue below that this is due to mixing rotation in the disk with the blueshifted cone of the radial outflow. The dispersion is high in a larger region to the SW of the nucleus. We argue below that this is due to mixing rotation of the disk with the innermost part of filament W1 which crosses the nucleus in projection but maintains a blueshifted velocity, i.e. the filament kinematics does not follow rotation in the potential of the black hole. In figure \ref{fig:pvwfm} we present pv diagrams for the W1 and W2 filaments; they are discussed in Sect. \ref{wfmfil}.

\begin{figure*}
\centering
\includegraphics[scale=0.53,clip,angle=0]{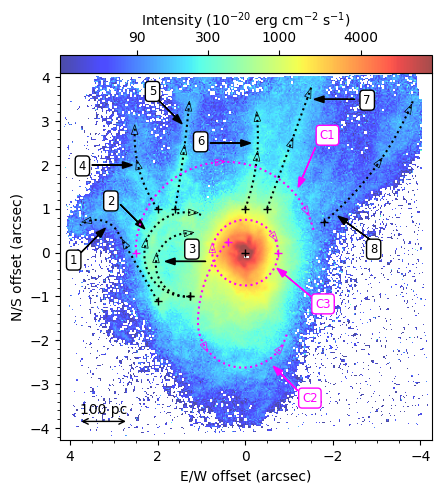}
\includegraphics[scale=0.53,clip,angle=0]{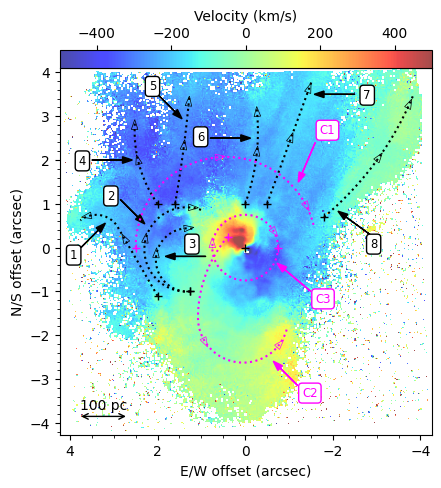}
\includegraphics[scale=0.53,clip,angle=0]{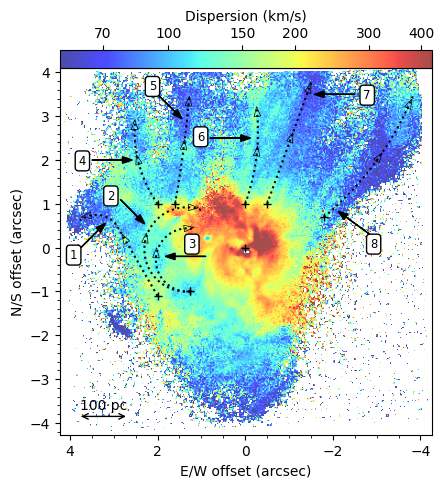}
\caption{\niitwo\ emission line moment maps from Gaussian fits to the \nfm\ cube. Left to right are maps of the total intensity, velocity, and velocity dispersion. Eight ionized gas filaments (black dotted lines), and three pseudo-slits (magenta dotted lines), are numbered and indicated with arrows of the same color in all panels.}
\label{fig:nfmmoms}
\end{figure*}

\begin{figure*}
\centering
\includegraphics[scale=0.55,clip,angle=0]{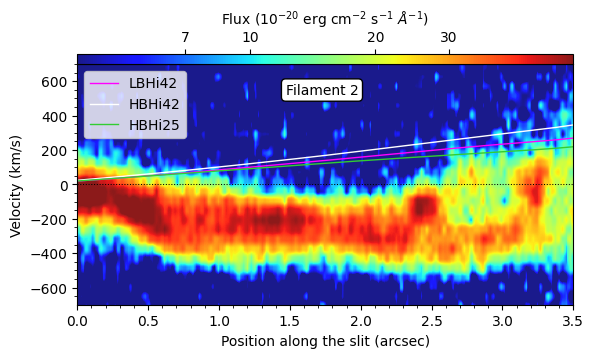} 
\includegraphics[scale=0.55,clip,angle=0]{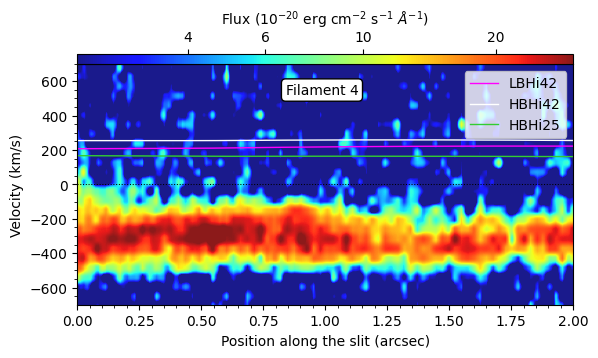} 
\includegraphics[scale=0.55,clip,angle=0]{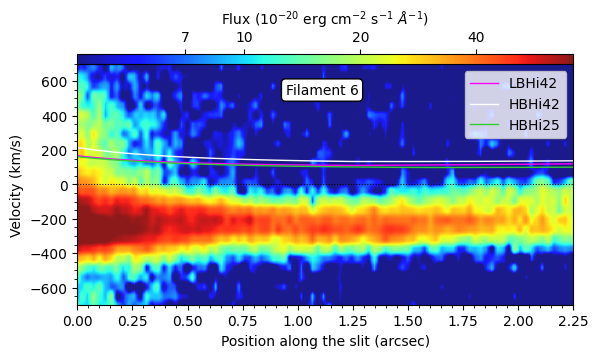} 
\includegraphics[scale=0.55,clip,angle=0]{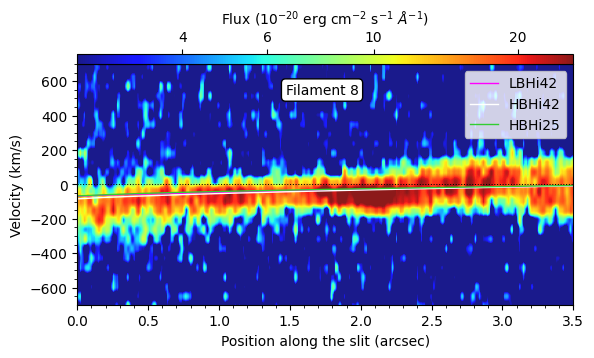} 
\caption{As in Fig. \ref{fig:pvwfm} but for four of the eight filaments marked with black dotted curves in the \nfm\ moment maps of Fig. \ref{fig:nfmmoms}. The remaining four filaments are presented in Fig. \ref{fig:apppvnfm}. These pv diagrams are discussed in Sect. \ref{nfmfil}, but are placed here for easy comparison with Fig. \ref{fig:nfmmoms}.}
\label{fig:pvnfm}
\end{figure*}

\begin{figure*}
\centering
\includegraphics[scale=0.55,clip,angle=0]{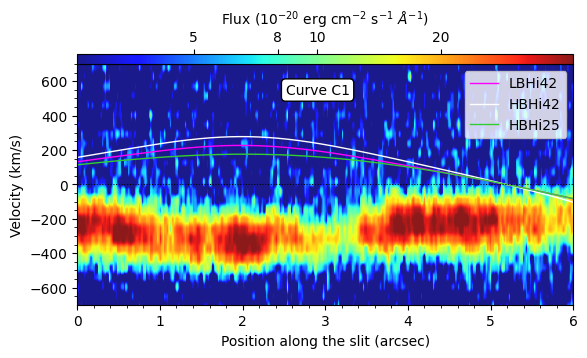}
\includegraphics[scale=0.55,clip,angle=0]{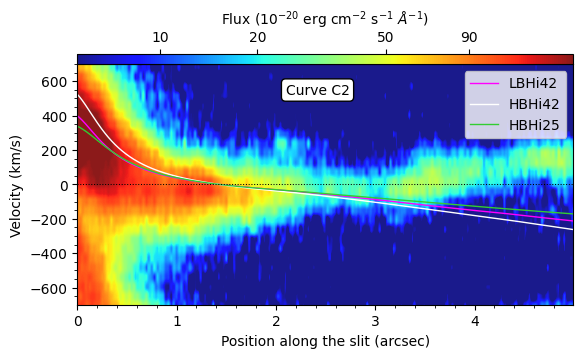}
\caption{As in Fig. \ref{fig:pvwfm} but for pseudo-slits C1 and C2 marked in magenta dotted curves in the moment maps of Fig. \ref{fig:nfmmoms}.
}
\label{fig:pvcurves}
\end{figure*}

\begin{figure*}
\centering
\includegraphics[scale=0.55,clip,angle=0]{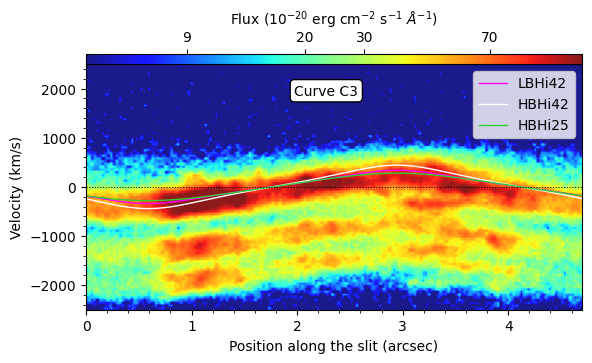}
\includegraphics[scale=0.55,clip,angle=0]{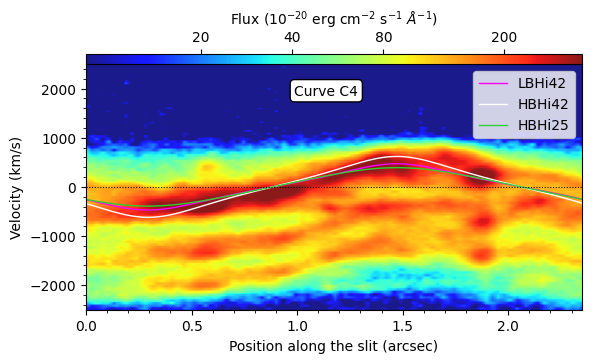}
\caption{Extra position-velocity diagrams of circular pseudo-slits in the \niione\ moment maps (Fig. \ref{fig:nfmmoms})
Left: as in Fig. \ref{fig:pvwfm} but for the pseudo-slit C3 (nuclear radius 0\farcs75) marked with a magenta dotted line in the left and central panels of Fig. \ref{fig:nfmmoms}. Right: as in the left panel, but for a pseudo-slit at a nuclear radius of 0\farcs375. The range of velocities is larger in order to show the full \hanii\ line complex.} 
\label{fig:pvcurves2}
\end{figure*}

\subsection{\ha\ + \nii\ \nfm\ Moment maps}
\label{secmomwfmha}

Equivalent \nfm\ moment maps are presented in Fig. \ref{fig:nfmmoms}, with the locations of several pseudo-slits identified. The black dotted lines correspond to eight possible ionized gas filaments: most are easily distinguishable in at least one of the three moment maps. The three pseudo-slits shown with magenta dotted lines do not denote filaments; these pseudo-slits cross regions in the map whose velocity features help us better understand the nuclear kinematics. The nuclear patterns seen in the \wfm\ moment maps (Fig. \ref{fig:wfmmoms}) - the blue region to the N, the posited rotating disk, and the multiple filaments - are now resolved in exquisite detail. Given the rich complexity of the components and kinematics, these \nfm\ moment maps will be more extensively discussed in Sects. \ref{nfmfil} and \ref{discussion}. At this point we note the diffuse and constant velocity blue- (red-) shifted emission to the N (S) of the posited rotating accretion disk which have high velocities but low dispersion, the multiple filaments to the N (with velocities apparently lower than the diffuse blue emission), and the high dispersion in the posited accretion disk. The latter is caused, in part, from the mixing of multiple velocity components: the disk and outflow (N), and the disk and a filament crossing the nucleus in projection (S). Figures \ref{fig:pvnfm}, \ref{fig:pvcurves} and \ref{fig:pvcurves2} show pv diagrams for some of the filaments; they are discussed in Sect. \ref{nfmfil}.

\begin{figure}
\centering
\includegraphics[scale=0.70,clip,angle=0]{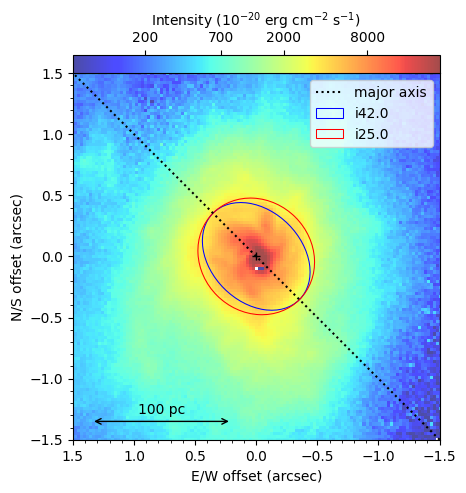} 
\caption{Zoom into the \nfm\ moment 0 image of the \niitwo\ line, to better illustrate the spiral arm pattern and disk morphology. For reference, the blue (red) ellipses show the expected morphology of a thin disk with inclination
42\arcdeg\ (25\arcdeg), the values favored by W13 and this work, respectively}
\label{fig:niiell}
\end{figure}

A detailed view of the inner ionized disk as traced by the \niitwo\ total intensity is shown in Fig. \ref{fig:niiell}.
\citet{for94} used isophotal fits to the inner arcsec ionized disk (which we show in this work to be a mix of different components, not only a rotating disk) to derive a PA of 1\arcdeg--9\arcdeg\ (significantly different from the kinematic PA of $\sim$40--42\arcdeg\ supported by all kinematic studies which allowed this PA to vary) and an inclination of 42\arcdeg. 
Here, complex filamentary structure is still notable, and we thus concentrate only on the spiral arms. The most plausible explanation for these spiral arms are trailing arms in a rotating disk as noted by \citet{for94}. They are thus the most reliable indicator of the true extent of the disk, and thus its inclination. As can be seen in Fig. \ref{fig:niiell}, the outer ends of the spiral arms are visually better enclosed in an i = 25\arcdeg\ (or even lower), rather than i = 42\arcdeg, inclined disk.
 
\subsection{\oi\ \nfm\ Moment maps}
\label{secmomnfmoi}

In the innermost arcsecond, the large intrinsic linewidths and the presence of multiple velocity components (especially highly blueshifted components) make the  automated Gaussian fits to the \hanii\ lines relatively unreliable (though these multiple components are still visually distinguishable in pv diagrams). Among the other emission lines in the \nfm\ cube, our best option is the \oione\ line as it is relatively isolated and has sufficient SNR.  

\begin{figure*}
\centering
\includegraphics[scale=0.5,clip,angle=0]{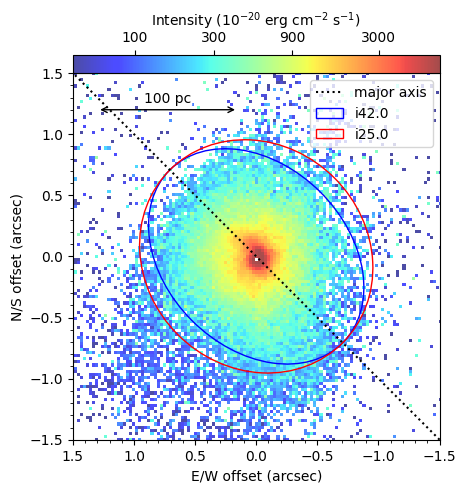} 
\includegraphics[scale=0.5,clip,angle=0]{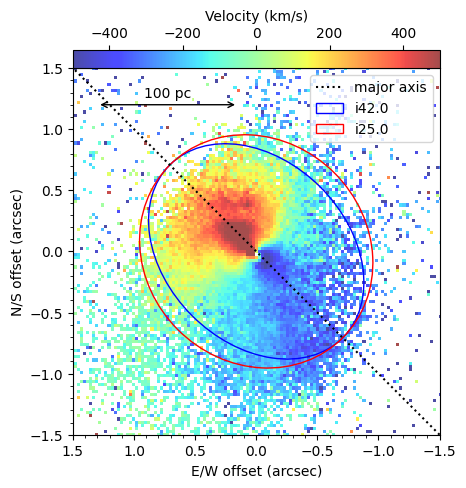}
\includegraphics[scale=0.5,clip,angle=0]{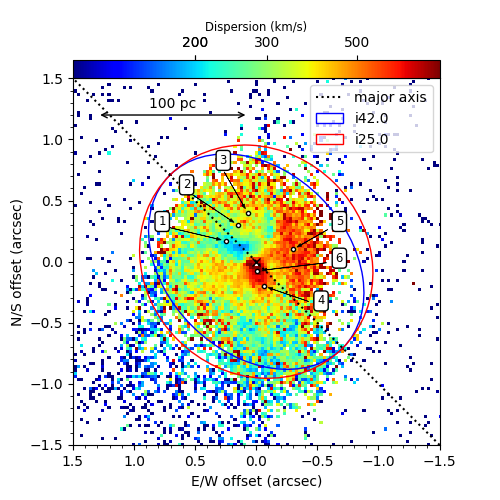}
\includegraphics[scale=0.5,clip,angle=0]{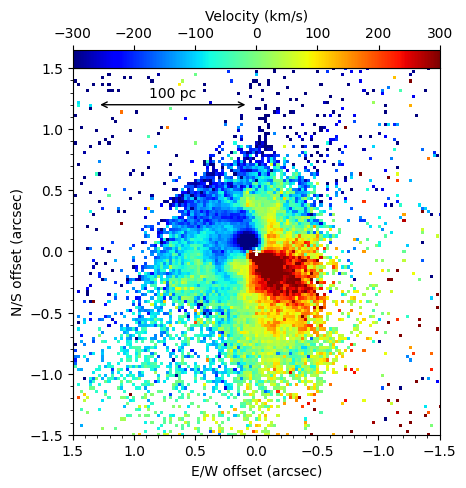}
\includegraphics[scale=0.5,clip,angle=0]{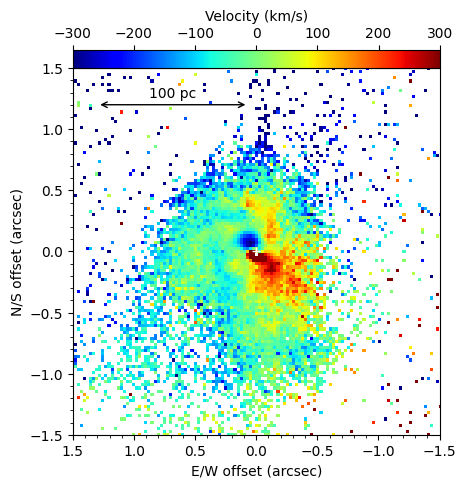}
\includegraphics[scale=0.5,clip,angle=0]{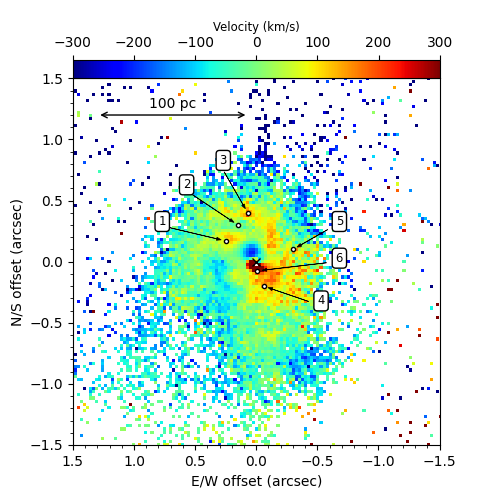}
\caption{As in Fig. \ref{fig:nfmmoms} but for the \fulloi\ line in the MUSE \nfm\ cube. Only a subset of the \nfm\ FOV, in which the SNR of the \oi\ line is high enough, is shown. For illustration, the blue (red) ellipses in the top panels show the expected morphology of a thin disk with inclination 42\arcdeg\ (25\arcdeg), the inclination values favored by W13 and this work, respectively. In the rightmost panels, a cross marks the kinematic center; the apertures marked with black circles and labeled with numbers 1 to 6, are used in Fig. \ref{fig:spectra}.}
\label{fig:oimoms}
\end{figure*}

\begin{figure*}
\includegraphics[scale=0.28,clip,angle=0]{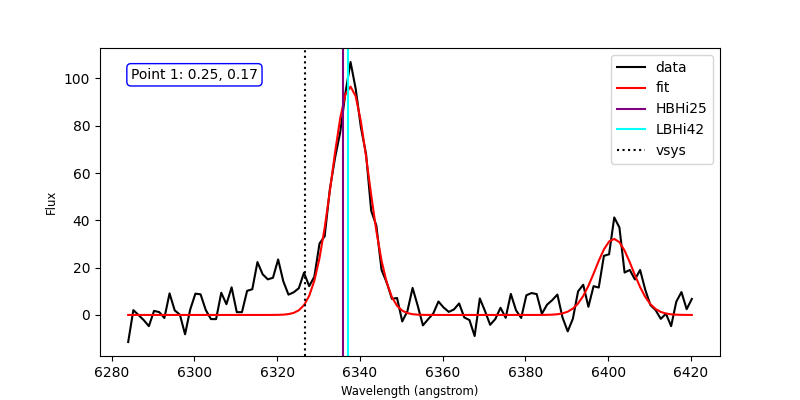}
\includegraphics[scale=0.28,clip,angle=0]{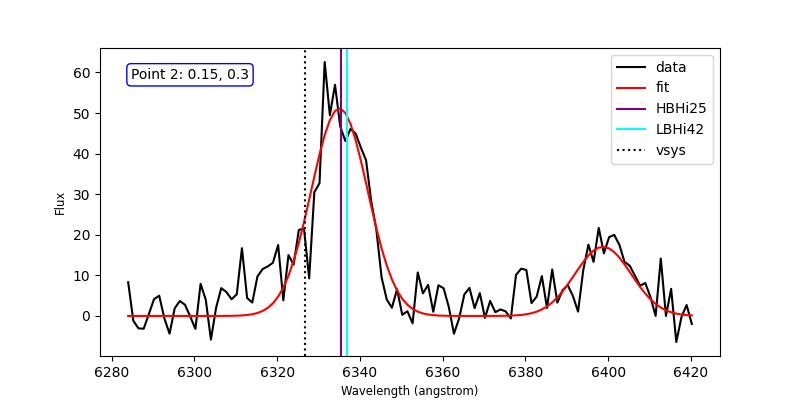}
\includegraphics[scale=0.28,clip,angle=0]{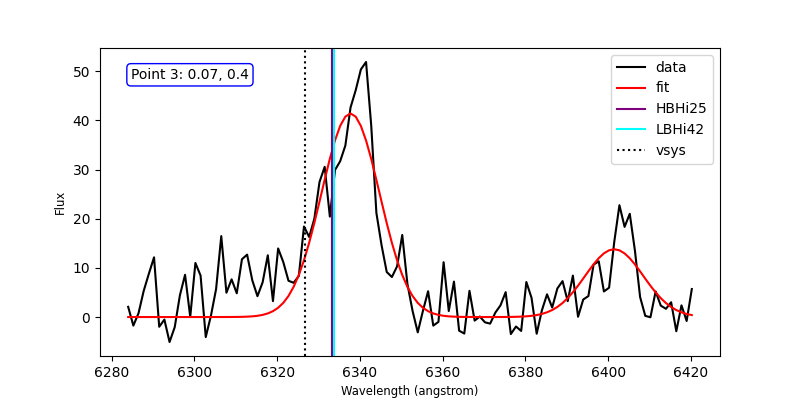}
\includegraphics[scale=0.28,clip,angle=0]{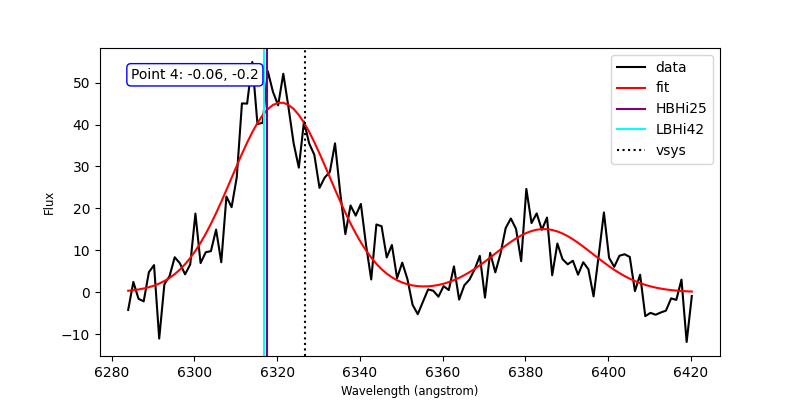}
\includegraphics[scale=0.28,clip,angle=0]{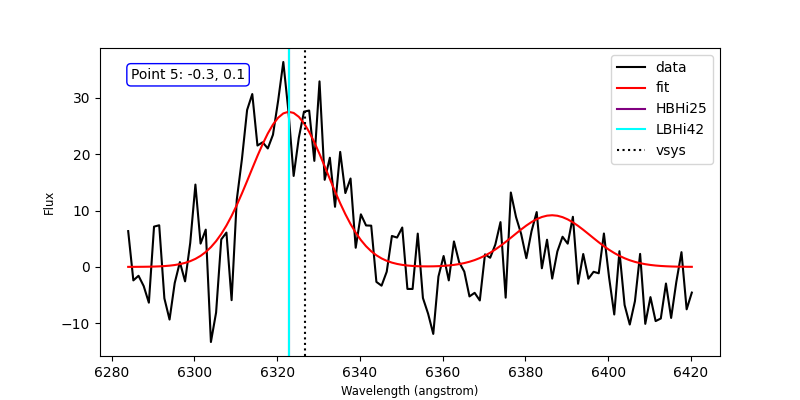}
\includegraphics[scale=0.28,clip,angle=0]{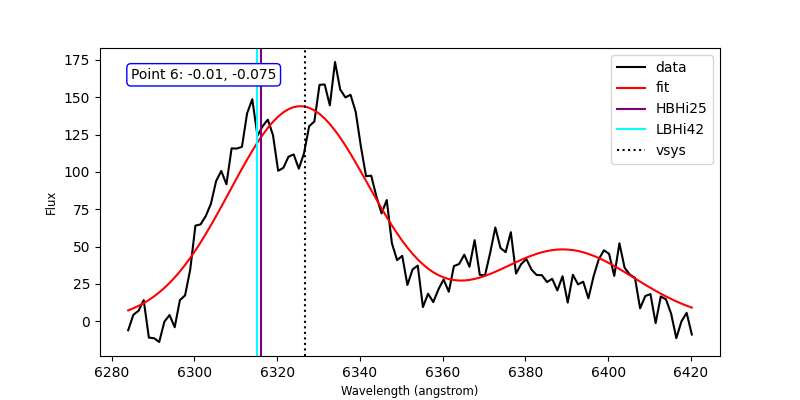}
\centering
\caption{\oi\ spectral profiles (black solid lines) in selected apertures of the nuclear ionized disk for the apertures indicated in the right panels of Fig. \ref{fig:oimoms}, with their single Gaussian fit overlaid in red; the strong line is \oione\ and the weaker line to the red is \oitwo. The systemic velocity (black dotted line), and rotation velocities in models with \mbh\ = 6.0 $\times\ 10^{9}$ \msun\ and i = 25\arcdeg\ (HBH i25; purple) and \mbh\ = 3.5 $\times\ 10^{9}$ \msun\ and i = 42\arcdeg\ (LBH i42; cyan) are overplotted. Apertures are labeled by their number and their RA and Dec offset from the kinematic center in arcsec. Apertures were selected to sample portions of the disk with different velocity dispersions and velocity residuals. Aperture 1 is located in the lowest dispersion area of the disk. 
Aperture 6 shows two components: one centered on the expectation of rotation and the other, with potentially larger flux, which is redshifted 950 \kms\ from the rotating component, and 330 \kms\ from systemic velocity. 
Apertures to the N (1, 2, 3) show smaller dispersions, plus a potential component blueshifted by $\sim$350 \kms\ which is much weaker in flux than the rotating component.
}
\label{fig:spectra}
\end{figure*}

Figure \ref{fig:oimoms} (top row) shows the moment maps for the \oione\ line in the \nfm\ data cube; the FOV is smaller here as we only show regions with relatively high SNR.  
The intensity (moment 0) map shows decreasing flux with increasing nuclear distance but does not clearly delineate an axisymmetric inclined disk with a clearly identifiable inclination. 
For reference we overplot, in blue and red  lines, illustrative ellipses expected for a circular disk with inclination 42\arcdeg\ (the value determined by W13) and 25\arcdeg\ (the value we favor in this work; see Sects. \ref{igkinemetry} and \ref{igkinms}), respectively.
The velocity map shows a clear rotational pattern, but with a twist close to the nucleus, and an amplitude which is not symmetric across the minor axis; the blueshifted emission has smaller amplitudes than the redshifted emission in the radius range 0\farcs1--0\farcs3.
In the next sections we argue that this is due to a filament which does not participate in rotation and crosses the disk in projection at a constant blueshifted velocity (which is not as blue as the expectation from rotation). 
The velocity dispersion is also non axisymmetric: 
In the highest SNR inner ($\leq$0\farcs2) disk, the blueshifted side of the disk has a larger velocity dispersion, reaching values of $\sim$600 \kms, to the SW. 
Spectra here are double-peaked with a red component at velocity similar to the redshifted side of the disk, and a blue rotating component (see Fig. \ref{fig:spectra}). 
The redshifted component could be attributable to the innermost region of the bi-conical outflow (Sect. \ref{igoutflow}) or a filament crossing the nucleus in projection (Sect. \ref{nfmfil}).
A high dispersion region 0\farcs5 to the W of the nucleus, in an area with a lower SNR, is also seen.
Finally, a region to the N has the lowest velocity dispersion observed ($\sim$170 \kms),
which likely traces the intrinsic dispersion of the disk (Fig. \ref{fig:spectra}).

The asymmetries in the dispersion and residual velocity maps of Fig. \ref{fig:oimoms} merit further details, especially given the filaments which cross the rotating disk in projection in the \nfm\ field of view. The right panels of Fig. \ref{fig:oimoms} (i.e. the dispersion and HBH i25 residual maps) show the locations of six apertures whose \oi\ spectra are shown in Fig. \ref{fig:spectra}. 
Apertures 1 and 3 are in the 'spiral arms' seen in the residual velocity maps, while aperture 2 is in a region between them. As seen in these three spectra, the central velocity of the best fit single Gaussian coincides with the peak velocity of the spectrum, confirming that the velocity map from Gaussian fitting reproduces well the central velocity of the strongest emission component. However, all three apertures show a potential weaker (and well separated) component near $\sim$6320\AA. This weak blueshifted component is at the expected velocity of our posited blue conical outflow to the NE (Sect. \ref{igoutflow}). 
Aperture 1 is in the region of the lowest velocity dispersion of the nuclear ionized disk. It is cleanly, and at high SNR, fit with a single Gaussian with dispersion $\sim$170 \kms. This dispersion likely reflects the true dispersion of the disk in the absence of additional non-rotating velocity components. 
Apertures 4, 5, and 6 are in regions of high velocity dispersion. Aperture 4 seems to have a single component with high dispersion and a redshifted velocity with respect to the predictions of both HBH i25 and LBH i42 models; it is possible that they are a blend of two components with approximately the same intensity. The double component nature is most clearly seen in the spectrum of aperture 6: here a rotating component is distinguishable from a redshifted component: 950 \kms\ from the rotating component and $\sim$350 \kms\ from systemic; this velocity offset would be expected if the gas were in the red cone of the nuclear outflow (Sect. \ref{igoutflow}).
Aperture 5 (W of the nucleus) also has high velocity dispersion, but the SNR here makes it impossible to tell if it is formed from one or more components. Overall, apart from small velocity offsets from models, regions of the disk show higher velocity dispersion due to additional components, and should be  masked when comparing observed velocities to simulated rotating model cubes. Specifically, we mask the high velocity dispersion regions and/or the spiral arm feature regions in Sect. \ref{rvmaps}.

\section{Ionized disk geometry: PA and Inclination from \kinemetry}
\label{igkinemetry}

\begin{figure*}
\includegraphics[scale=0.5,clip,angle=0]{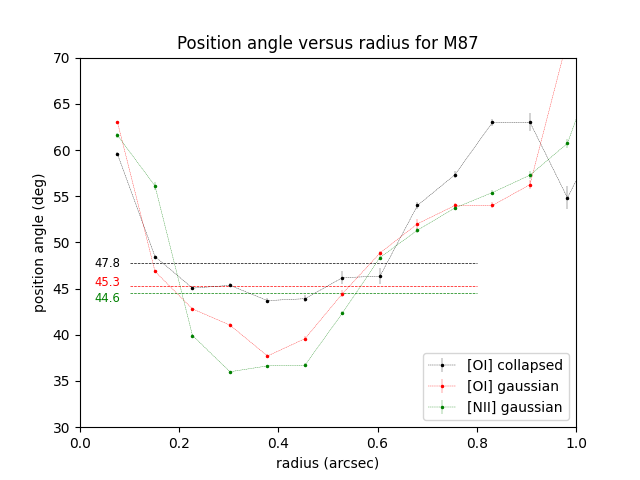}
\includegraphics[scale=0.5,clip,angle=0]{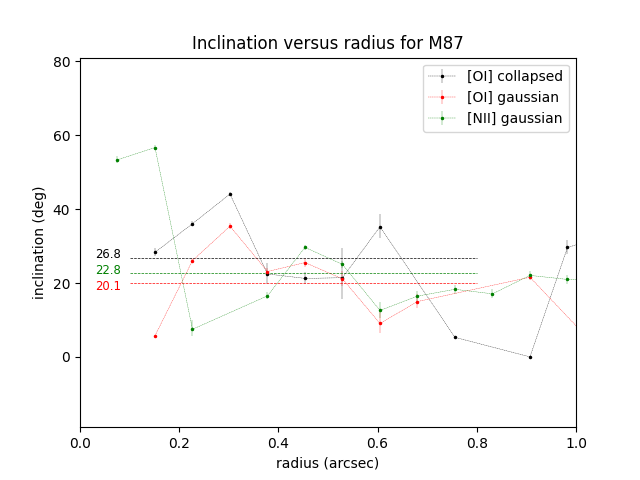}
\includegraphics[scale=0.5,clip,angle=0]{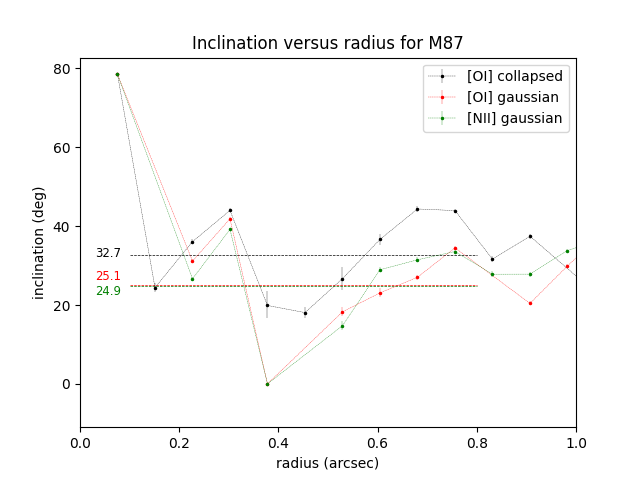}
\includegraphics[scale=0.475,clip,angle=0]{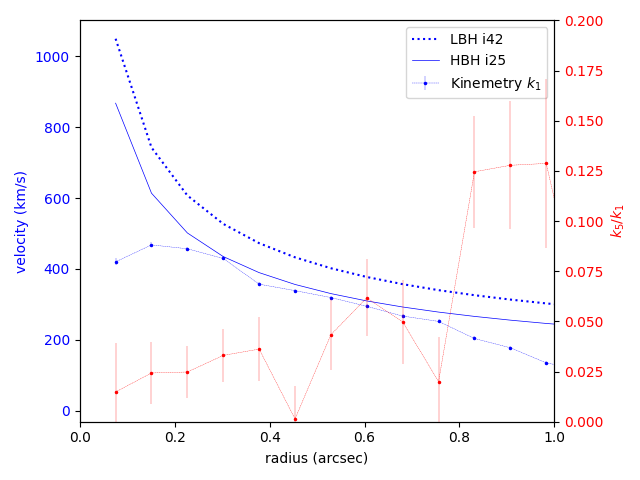}
\centering
\caption{Results from \kinemetry\ for the sub-arcsec ionized gas velocity maps: the top two panels show results when both disk PA and inclination are allowed to vary, while the bottom two panels show results when the PA is fixed at 45$\dg$ and only inclination is allowed to vary. Specifically, the top panels show the variation of the best fit PA (left) and inclination (right) with nuclear radius. The bottom left panel shows the variation of the inclination when the PA is fixed to 45$\dg$. The velocity maps used were the single Gaussian fit velocity maps of \oi\ (red) and \ha\ + \nii\ (green) and the 'collapsed' moment 1 map of the \oi\ line (black).
In each of these three panels the horizontal lines in the corresponding color show the mean value over radii between between 0\farcs1 and 0\farcs8. 
The bottom right panel shows the rotation curve (i.e. the '$k_{1}$' parameter from \kinemetry; connected blue dots) derived by \kinemetry\ for the Gaussian-fit \oi\ velocity map, and compares this to expectations of Keplerian rotation for the HBH i25 (solid blue line) and LBH i42 (dotted blue line) models, all following the left hand $y$ axis. The red line, following the right hand $y$ axis shows the $k_{5}/k_{1}$ ratio (i.e. non-circular velocity to circular velocity ratio) derived by \kinemetry: low values here imply that disk velocities are dominated by rotation. At very small radii too few pixels are included in each ring and \kinemetry\ results are unreliable.
}
\label{fig:pakinemetry}
\end{figure*}

In this section we determine the best fit PA and inclination of the ionized gas disk using \kinemetry\ (see Sect. \ref{metandsoft}). Three different velocity maps were independently input to \kinemetry: the velocity maps from the single Gaussian fitting of \oi\ and of \ha\ + \nii, and also a directly 'collapsed' \oi\ velocity map.
The directly 'collapsed' velocity map is a moment 1 map of the \oione\ line, created with the \textit{spectral-cube} package in python, which only included a limited wavelength (thus velocity range); to avoid contamination from the weaker \oitwo\ line (63 \AA\ or 3000 \kms\ redward in restframe), at each spaxel we only included a range of $\pm$40\AA\ (or $\sim\pm$1900\kms) centered on the wavelength predicted by the HBH i25 rotational model for that spaxel. 

In the first run, for each velocity map, both  PA and inclination were free parameters: the best-fit results are shown in the top two panels of Fig. \ref{fig:pakinemetry}. In the case of PA, 
all three input velocity maps give relatively similar results: the PA starts at value between 58\arcdeg\ and 64\arcdeg\ at 0\farcs1, decreases out to radius 0\farcs4, and increases beyond this radius. Our kinematic analysis further below concentrates on radii between  0\farcs1 to 0\farcs8: in the inner 0\farcs1 the lines become broad and blended and velocity isophotes are twisted; farther than 0\farcs8 the SNR of disk ionized gas is low and other components (outflow and filaments) dominate. 
While only the 'collapsed' \oi\ velocity field shows a PA more or less constant over a large part of this useful radius range, the mean PA over this radius range is close to 45\arcdeg\ for all three velocity fields.
The top right panel of Fig. \ref{fig:pakinemetry} show the equivalent results for the best fit inclination. None of the velocity fields give a constant inclination with radius, and the inclination evolution appears to be relatively chaotic with values $\sim$0--45\arcdeg. The mean inclination values between 0\farcs1 and 0\farcs8 shown are between 17\arcdeg\ and 24\arcdeg\ for the three velocity maps. 

Given the results of the first run of \kinemetry\ (and the value previously derived by W13), we decided to set a fixed PA of 45\arcdeg\ and leave only the inclination as a free parameter in the second run of \kinemetry. The bottom left panel of Fig. \ref{fig:pakinemetry} shows the result of these runs. 
The variation of inclination with radius is now slightly less chaotic than in the first run but still varies between 20\arcdeg\ and 45\arcdeg, when it is not 0\arcdeg. The mean values over the 0\farcs1--0\farcs8 range are remain between 17\arcdeg\ and 24\arcdeg, except for the 'collapsed' \oi\  map, which prefers a higher mean value of 33\arcdeg. 

While these \kinemetry\ results do not concretely point to a disk inclination close to 25$\dg$, they provide an additional indication that inclination values closer to 20$\dg$--25$\dg$ are more plausible than 42$\dg$. It is also an indication  of the relatively complex gas motions in the disk could preclude an accurate measurement of the black hole mass. 

\section{Ionized gas filaments}
\label{igfilaments}

In this subsection we present and discuss the kinematics of the ionized gas filaments seen in the outer (\wfm) and nuclear (\nfm) regions. We present evidence that each has its own flow velocity, and only some are affected by the SMBH potential, while others cross the nucleus only in projection. That is, the nuclear 'gas disk' is not an isolated relaxed and well-ordered rotating disk, but a mix of gas rotating in a 'disk', emission from gas in filaments, some of which enter with their own proper bulk velocity, and some of which only cross the nucleus in projection. Further below, we  also discuss the extra complexity of disentangling the nuclear biconical outflow.

\subsection{Wide Field Mode filaments}
\label{wfmfil}

In Sect. \ref{wfmfil} we highlighted two large-scale filaments in the \wfm\ moment maps (Fig. \ref{fig:wfmmoms}). The pv diagrams of the \nii\ emission line along these filaments are shown in Fig. \ref{fig:pvwfm}, together with the predictions of disk rotation in the potential of the SMBH and the galaxy under different assumptions of black hole mass and disk inclination. 
Since the apertures are curved, we refer to these as pseudo-slits.
The offsets ($x$-axis) in the pv diagrams start at the position of the nucleus, and continue along the dashed lines shown in Fig. \ref{fig:wfmmoms}. That is, the $x$-axis value indicates the distance traveled along the line from the nucleus (indicated with a black cross in Fig. \ref{fig:pvwfm}) and not the radial distance from the nucleus. The colored solid lines in the pv diagrams denote the velocities expected from gas in a rotating thin disk for the following models: LBH i42 (magenta), HBH i42 (white), and HBH i25 (green).

All of these models have a disk in PA = 45$\dg$. The galaxy potential is obtained as described in Sect. \ref{metandsoft}, and becomes significant only at radii beyond $\sim$2\arcsec. The LBH i42 and HBH i42 (magenta and white) models use the inclination and position angle found by W13 for the ionized gas disk. The green line is illustrative for a HBH and a disk with the lower inclination suggested by our data (see Sect. \ref{igpvs}).

Filament W1 is the longest and most prominent. None of our velocity models follow the observed velocities in the first 14\arcsec. The filament leaves the nucleus in PA roughly close to the minor axis of the W13-posited gas disk and one would expect relatively small Keplerian velocities. Nevertheless, velocities are between $\pm$100 \kms\ (with an extreme value of $\sim$400 \kms) over the first $\sim$30\arcsec\ of the filament. 
Between 20\arcsec\ and 30\arcsec\ there are two - potentially bubble-shaped - clouds at 90 \kms\ and $-$350 \kms. The latter has a velocity highly different from the rest of the filament and our models; this bubble is either not part of the filament (but in the same line of sight) or caused by a kink or instability in the filament.
From 36\arcsec\ onward, the observed velocities follow the overall pattern expected from the HBH i25 rotating disk model in both values and shape. To make the HBH i42 and LBH i42 models agree with the observed velocities over this range, one would require to change the ionized gas 'zero' velocity by $\sim$80 \kms\ to the blue, which would imply an even greater difference between the recessional velocities of ionized gas and stars ($\sim$1300 \kms). 

Filament W2 is shorter and  shows relatively smooth velocity changes along its length, with a relatively low dispersion (compared to filament W1) at offsets larger than 3\arcsec.
Similar to filament W1, 
the observed velocities close to the nucleus are systematically bluer than all our rotating disk models, especially in the first 5\arcsec. Once more, at distances larger than 10\arcsec\ the HBH i25 model most closely fits the data in values and shape. 
It is also interesting to note that filament 2 appears to (at least in projection) connect to filament W1 through an additional highly redshifted bridge to the E of the nucleus (Fig. \ref{fig:wfmmoms}), though we find no  explanation for this velocity pattern.

Except in the innermost $\sim$5\arcsec--12\arcsec, the observed pv diagrams of both filaments W1 and W2 are well explained - in velocity and velocity gradients - by circular rotation of gas in a disk with inclination close to 25\arcdeg\ (green line in the pv diagrams). This can also be seen in the residual velocity maps (bottom row of Fig. \ref{fig:wfmmoms}). At these large radii, the galaxy potential, rather than SMBH mass, drives the circular velocity. This is clearly illustrated by the fact that the HBH i42 (white lines) and LBH i42 (magenta lines) models result in very similar velocity predictions. The disk inclination is thus the dominant parameter, and the degeneracy between black hole mass and inclination when fitting the kinematics of the inner arcsec of the ionized gas disk no longer plagues us. 

\subsection{Narrow Field Mode filaments}
\label{nfmfil}

We now focus on the eight filaments and three pseudo-slits identified on the higher resolution \nfm\ moment maps (Fig. \ref{fig:nfmmoms}). The pv diagrams of the \niitwo\ line along these filaments and pseudo-slits are shown in Figs. \ref{fig:pvnfm}, \ref{fig:pvcurves}, and \ref{fig:apppvnfm}.
Recall that the $x$-axis of these figures are offsets along a curved slit, and not the radial distance from the nucleus. Most of the filaments are to the north of the nucleus, and cross the highly blueshifted region. The pseudo-slits (magenta lines in Fig. \ref{fig:nfmmoms}) do not follow gas filaments, but their pv diagrams are useful to understand the complex nuclear kinematics, so they are also presented here. 

Each of the eight filaments in Fig. \ref{fig:pvnfm} can be differentiated from the others in at least one of the three moment maps in Fig. \ref{fig:nfmmoms}. We expect to see significant effects of the black hole potential on the ionized gas as we get closer to the nucleus, especially in the inner few arcseconds. However, the filament pv diagrams (Figs. \ref{fig:pvnfm} and \ref{fig:apppvnfm}) are almost all significantly different from all of the model predictions of thin disk rotation in the black hole and galaxy potential. Only filament 8, which is to the W of the highly blueshifted region (see middle panel of Fig. \ref{fig:nfmmoms}) approximately follows our rotating disk model predictions. 

Filament 1 can be distinguished in the intensity map; its pv diagram (Fig. \ref{fig:apppvnfm}) shows differences of up to $\sim$400 \kms\ blueward from the rotation model predictions. 
Filaments 2 and 3 have a semicircular form in the moment maps, with their extremes closest to the nucleus. Both can be distinguished in the intensity map, with filament 2 also partially distinguishable in the velocity and velocity dispersion maps. For both filaments, the disagreement between observed velocities and the rotational models increases with increasing nuclear distance. The disagreement is nevertheless smaller close to the nucleus than the case of filament 1. 
Filaments 4 and 5, clearly distinguishable in the intensity maps, are in the highly blueshifted zone to the NE of the nucleus. They are the most blueshifted filaments in the inner nucleus, with blueshifts of $\sim$250--350 \kms\ from systemic, and more than 500 \kms\ offset from the rotation model predictions. Nevertheless, their dispersion velocities are $\leq$150 \kms\ and both appear to have only a single velocity component.
Filaments 6 and 7 are distinguishable in the intensity and velocity maps. With velocities between $-$200 \kms\ and $-$150 \kms, they are less blue shifted than filaments 4 and 5. Filament 6 shows a higher dispersion ($\sim$200 \kms) than filaments 4, 5 and 7. 
Filaments 4 to 7 in the \nfm\ can also be distinguished in the \wfm\ maps:  at the lower resolution of \wfm, they are distinguishable as three filaments (or one filament and a loop).

Filament 8, the westernmost filament, is clearly distinguished in the intensity and velocity map. It is the only filament to approximately follow our rotating disk models, though it is slightly more blueshifted than the models at offsets of $\sim$\ifas{1}{2} and $\sim$1\farcs8--2\farcs2. Its dispersion is also small ($\leq$150 \kms). Given its velocity and  morphology, it is likely the inner region of filament W1 of the \wfm\ (Fig. \ref{fig:wfmmoms}). 

Overall, filaments 1 to 7 follow individual flow velocities which cannot be explained by a rotating thin disk. Only filament 8 approximately follows the expectations of a rotating thin disk, but since this filament approaches the nucleus in a PA close to the minor axis of the disk, it cannot be used to constrain the black hole mass.

We now discuss the pv diagrams (Fig. \ref{fig:pvcurves}) of the pseudo-slits C1 to C3 in the \nfm\ cube (dotted magenta curves in Fig. \ref{fig:nfmmoms}). 
The pseudo-slit C1 (left panel of Fig. \ref{fig:pvcurves}) cuts through most of the northern filaments from east to west while maintaining a roughly constant radial distance from the nucleus. Nevertheless its pv diagram shows a constant blueshift of $\sim$200 \kms\ from systemic except over offsets of $\sim$\ias{1}{3} along the pseudo-slit (effectively the NE region) where a constant blueshift of $\sim$300 \kms\ is seen. While this is visible in the velocity map, the pv diagram emphasizes the relatively uniform velocity of the northern filaments, the uniformly small dispersion velocity, and the lack of additional velocity components. Effectively the filaments are bright concentrations embedded within, and sharing the velocity of the more diffuse blueshifted emission region. The observed velocities are even further blueshifted from the rotation model predictions, and in fact appear as a rough mirror image of these.  

The pseudo-slit C2 starts in the redshifted side of the inner disc, travels to the SE along a small filament (see left panel of Fig. \ref{fig:nfmmoms}) and then crosses the redshifted ionized gas to the south in the \nfm\ moment map. In its pv diagram (right panel of Fig. \ref{fig:pvcurves}), we note the high dispersion in the first arcsec, and the fact that the kinematics roughly follow the LBH i42 and HBH i25 rotating disk models in the inner two arcsec. We see a small velocity discontinuity at $\sim$2\farcs5--3\arcsec\ offset along the pseudo-slit,  which delineates the separation point between the rotating (inner) and non rotating (outer) components of the gas. However, the continuity of the velocity transition from one component to the other  indicates interaction between the two components.

The pseudo-slit C3 is a circle with radius 0\farcs75 centered on the nucleus. The $x$ axis starts (offset = 0) W of the nucleus and moves clockwise. Over all of the pseudo-slit except the NW quadrant the observed velocities are well fit with either the LBH i42 model, or a HBH i25 model. In the NW quadrant, the observed velocities are systematically $\sim$200 \kms\ blueshifted from the  model predictions. While filaments 6 and 7 of the \nfm\ can be seen as separate components at even larger blueshifts, the $\sim$200 \kms\ blueshift in the NW does not appear to be an additional component but rather a warp or systematic bulk velocity in the disk. 
At this radius, the S (especially the SW) section of the disk appears the least confused with additional components or velocity offsets. 
The pv diagram shown in the right panel of Fig. \ref{fig:pvcurves2} is for a circular slit like C3, but at half its radius (now r = 0\farcs375). Here, the disk velocities are more symmetric in all PAs, with the LBH i42 and the HBH i25 both providing a reasonable fit. We note that the N blueshifted filaments are still clearly distinguishable. At this smaller radius, the gas to the SW shows a second velocity component: a non-rotating component $\sim$300 \kms\ blueshifted from systemic velocity. This component dominates the rotating component at smaller radii, and is the primary reason why the blueshifted side of the rotating disk shows smaller absolute velocities than the redshifted side of the disk over radii 0\farcs1--0\farcs3. The velocity of this non-rotating component matches that of filaments 6 and 7 as they approach the nucleus in projection. 

\section{Ionized gas rotation: position-velocity diagrams and comparison to analytic models}
\label{igpvs}

\begin{figure*}
\centering
\sidecaption
\includegraphics[scale=0.45,clip,angle=0]{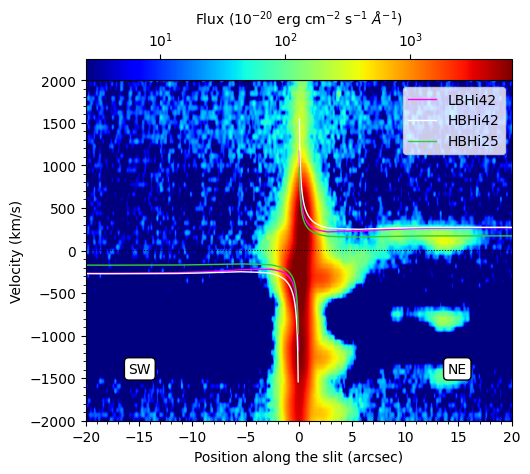}
\includegraphics[scale=0.45,clip,angle=0]{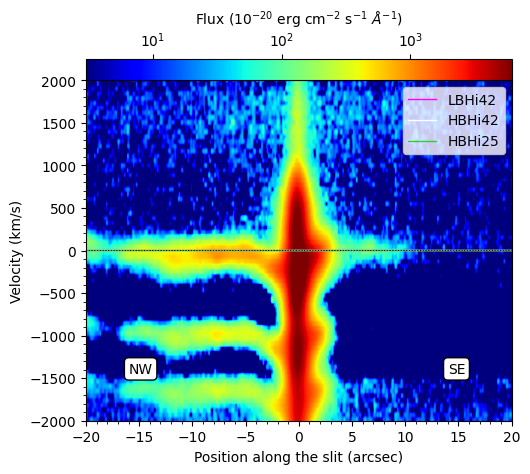}
\caption{PV diagrams of the \fullhanii\ lines in the \wfm\ cube along the W13 major (45\arcdeg, left) and minor axes (135\arcdeg, right). The models (see inset) and the 'zero' velocity are plotted with respect to the \niitwo\ line.}
\label{fig:wpvmajor}
\end{figure*}

\begin{figure*}
\centering
\includegraphics[scale=0.45,clip,angle=0]{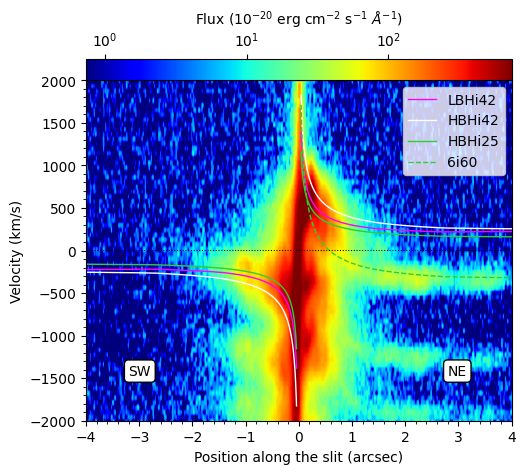}
\includegraphics[scale=0.45,clip,angle=0]{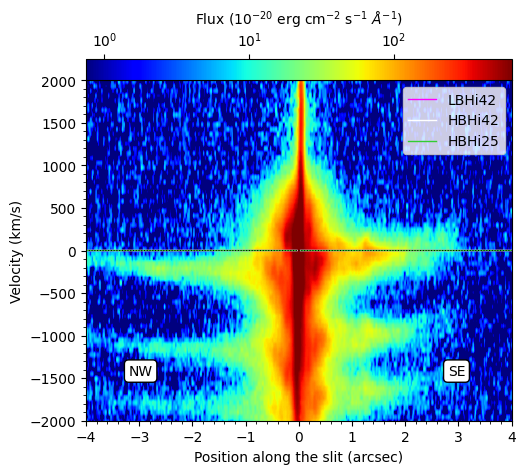}
\includegraphics[scale=0.45,clip,angle=0]{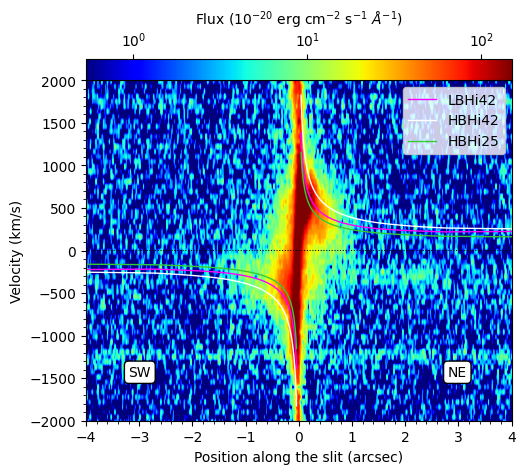}
\caption{PV diagrams of the \fullhanii\ (left and central) and \oi\ (right) emission lines in the \nfm\ cube, along the W13 major (45$\dg$, left and right) and minor axes (135$\dg$, central). Rotation models overplotted follow the inset. The dotted green line is for an illustrative rotating model with \mbh\ = 6.0 $\times$ 10$^9$ \msun\ (HBH), an inclination of 60\arcdeg, and a constant blueshift from systemic of 650 \kms.}
\label{fig:npvmajor}
\end{figure*}

Figure \ref{fig:wpvmajor} shows the pv diagrams of the \fullhanii\ line along nuclear slits in PA = 45\arcdeg\ (the W13-based disk major axis) and 135\arcdeg\ (disk minor axis) in the \wfm\ cube. 
Even at this relatively low spatial resolution, and at large nuclear offsets, the complex kinematics and presence of multiple components is already obvious. The inner 3\arcsec--4\arcsec\ to the NE shows highly blueshifted (instead of redshifted) emission, which, with increasing radius smoothly transitions to systemic velocity. Further, at 8\arcsec--16\arcsec, the bright emission line gas does not follow rotation in a disk with 42\arcdeg\ inclination. Along the minor axis to the NW, the nuclear gas is also blueshifted, and other kinematic 'wiggles' are seen further away from the nucleus.
Figure \ref{fig:npvmajor} shows the equivalent pv diagrams in the \nfm\ cube (left and middle panels) together with the pv diagram of the \fulloi\ line along a nuclear slit in PA = 45\arcdeg. In the left panel, the gas $\sim$1\farcs5--4\arcsec\ to the NE of the nucleus (positive offsets in the figure) primarily traces the kinematics of the filaments and outflow, and there is no clear sign of rotating gas at these offsets.  When the nuclear offset decreases below 1\farcs5, instead of continuing 'into' (in projection) the nucleus at the same bulk velocity offset, the gas velocity gets redder very rapidly but smoothly  (crossing systemic velocity at 0\farcs8 in projection) and then further increases until it is indistinguishable from gas in the rotating disk. 
Adding a simple bulk (systemic) blueshift to our rotation models cannot reproduce this velocity pattern (the observed velocities drop too rapidly with increasing nuclear distance). A filament crossing the nucleus only in projection would also not explain this pattern, as the projected nuclear separation is always larger than the true 3 dimensional nuclear distance. For illustration the dotted green line shows the predictions of gas rotating around a G11 black hole, but at an inclination of 60\arcdeg\ and with a bulk blueshift of 650 \kms. A possible explanation for this feature is that nuclear gas, initially feeding the black hole in a plane with a larger inclination than the main disk, is entrained into the blue cone of the conical outflow (see Sect. \ref{igoutflow}). 

Comparing the observed pv data to the models, the HBH i42 overpredicts the observed velocities, while both the LBH i42 and the HBH i25 models are relatively indistinguishable when compared to the data.  
Along the minor axis (central panel of Fig.  \ref{fig:npvmajor}), the blueshifted gas to the NW is clearly visible at an almost constant $\sim$200 \kms\ blueshifted velocity. 
The \oi\ pv diagram (right panel of Fig.  \ref{fig:npvmajor}) 
covers a smaller range of offsets. Here the HBH i42 model is clearly in the upper envelope of the observed velocities. 

All \wfm\ and \nfm\ pv diagrams clearly show filaments nearing the nucleus (in projection) but not following rotational kinematics of the black hole. In several cases, multiple velocity components are seen, so that single Gaussian fits are not usable here. While the highly blueshifted components can be disentangled in pv diagrams at most PAs, for reasons of brevity we cannot show all these here. To avoid some of the confusion created by the highly blueshifted filaments, we instead use the \oione\ emission line, which, while weaker, is relatively isolated from other emission and absorption lines, and is detected out to $\sim$1\arcsec\ radius (right panel of Fig. \ref{fig:npvmajor}). 

\section{Residual Velocity Maps}
\label{rvmaps}

The results of our \kinemetry\ (Sect. \ref{igkinemetry}) and residual (observed minus simulated) velocity map analysis (Sect. \ref{igkinms}) support low inclinations for the ionized gas disk. To better visualize the differences when using low and high inclination disks, and to discern residual velocity features after rotation velocity subtraction,  this section presents the residual (observed minus model) velocity maps of competing rotational models. The model velocity maps used in this section are created using \kinms.

It is important to note that the RIAF model includes a radial outflow in the disk plane, which results in an observed kinematic PA which is offset from the rotational kinematic PA. For this specific RIAF used here, a rotation with major axis in PA = 27$\dg$, added to the outflow component, results in an overall observed kinematic PA of  45$\dg$. Alternatively, if the near and far side of the disk were exchanged, then a rotation PA of 63$\dg$ would give the same result.

The bottom panels of Fig. \ref{fig:wfmmoms} show the residual velocity maps of the \niitwo\ line after subtracting the LBH i42, HBH i42, and HBH i25 models. At the relatively large nuclear distances seen in the \wfm, the galaxy potential dominates the black hole potential, thus avoiding the degeneracy between inclination and black hole mass in the inner arcsec. Both the HBH i42 and LBH i42 models (i.e. both models with i = 42\arcdeg) oversubtract the velocities seen over most of the filaments W1 and W2. The i = 25\arcdeg\ model however produces a velocity residual closer to zero.
Here it is important to note that we use a recession velocity of 1260 \kms, which is the velocity which best fits the inner ionized gas disk, and which is 75 \kms\ bluer than the value used by W13.
With our recessional velocity of 1260 \kms, the observed velocity fields of filament W2 (and a large part of filament W1) of the \wfm\ are primarily at velocities between systemic and 180 \kms\ (redshifted).
As discussed in Sect. \ref{wfmfil},  reducing the velocity residuals of the i = 42\arcdeg\ models requires using a smaller recessional velocity, which would be even further away from the stellar recessional velocity, and thus more unlikely to be correct.  
In any case, except for some prominent kinks and bubbles in filament W1, and extreme redshifts in the apparent bridge between the W1 and W2 filaments, both W1 and W2 appear to be primarily composed of gas in a rotating (and partially filled) ionized gas disk. The large blue region 1\arcsec--10\arcsec\ N of the nucleus, and the smaller red region $\sim$2\arcsec\ S of the nucleus, seen in the residual maps can be explained by the bi-conical outflow (Sect. \ref{igoutflow}).

Residual velocity maps for the \oi\ line are shown in the bottom row of Fig. \ref{fig:oimoms}: left to right are the residuals after subtracting the HBH i42, LBH i42 and HBH i25 models. The HBH i42 significantly over predicts the LOS observed velocity field. The LBH i42 and HBH i25 models give roughly similar residuals, though the latter provides velocity residuals closest to zero.

The twisted ionized gas velocity isophotes in the innermost 0\farcs1 (most clearly discerned in Fig. \ref{fig:oimoms}, but also clear on close inspection of Fig. \ref{fig:nfmmoms}) is an important result of this work. At a radius of 0\farcs3, the PA of the disk appears to be $\sim$45\arcdeg\ as found by W13 and \kinemetry\ (see Sect. \ref{igkinemetry}). By $\sim$0\farcs1 the PA appears closer to 90\arcdeg. In the innermost pixels the PA appears to twist even more. This innermost twist is best visualized in the residual velocity map (bottom right panel of Fig. \ref{fig:oimoms}). Here the residual map implies a rotation major axis of $\sim$18\arcdeg, which is perpendicular to the PA of the inner jet of M87, as would be expected if the jet was launched perpendicular to the accretion disk. Alternatively, this relatively N-S inner velocity gradient could trace the bases of the bi-conical outflow discussed in the Sect. \ref{igoutflow}. \citet{mac97} previously noted the velocity field twist inside 0\farcs2 when modeling their HST spectra with thin disk models. 
The twist in the apparent PA appears to continue outside the radius of 0\farcs3, as supported by the results in Sect. \ref{igkinemetry}.  
However, at even larger radii, as shown earlier in this section, the velocity field of the filaments in the \wfm\ maps at r $\sim$10s of arcseconds can still be fit well with a rotating disk in PA $\sim$45$\dg$ and inclination 25$\dg$.

\citet{ems14} have shown that the stellar velocity field in the central $\sim$15\arcsec\ of M87 is twisted, with the stellar kinematic PA changing rapidly from 17\arcdeg\ to $-$124\arcdeg. In the innermost $\sim$5\arcsec, the stars rotate with a major axis close to PA = 20\arcdeg, with the NE side redshifted. This pattern can also be seen in our \wfm\ stellar map (Fig. \ref{fig:wfmmoms}), where visually the inner $\sim$10\arcsec\ stellar kinematics support a PA of \idg{30}{45}, with the NE receding, i.e., similar to the pattern seen in the ionized gas here. The main difference is that the stellar rotation velocity amplitudes are relatively low ($\lesssim$20 \kms) and that the twisted isophotes are seen on much larger scales. 

The other important feature to note in the HBH i25 residual velocity map of Fig. \ref{fig:oimoms} is a three or four armed spiral pattern at $\sim$100 \kms\ redshift, embedded in a blue (outflow or filaments)
halo. This velocity-traced spiral arm pattern closely matches the spiral arm pattern noted by \citet{for94} in their HST ionized gas maps, and seen clearly in our moment 0 maps (e.g., Fig. \ref{fig:niiell}). The spiral pattern is reminiscent of the ionized gas spirals around SgrA*, which extend out to $\sim$3 pc, and appear to trace Keplerian circular or elliptical orbits around SgrA* \citep{Zhao2009}.      

\section{Observed vs. \kinms\ models of gas rotation: constraining black hole mass and inclination}
\label{igkinms}

To obtain a best fit model (in black hole mass and disk inclination) to the nuclear 'Keplerian' rotating ionized gas disk seen in the \nfm\ cubes, we use \kinms\ to construct simulated datacubes of the \oi\ line in the inner 1\arcsec,  
for a range of black hole masses (2 $\times\ 10^{9}$ \msun\ to 8 $\times\ 10^{9}$ \msun\ in steps of 0.5 $\times\ 10^{9}$ \msun) and disk inclinations (20\arcdeg\ to 50\arcdeg\ in steps of 2\arcdeg). In all models, the disk PA was fixed at 45\arcdeg, and the same galaxy potential (see Sect. \ref{metandsoft}) was used.

As described in Sect. \ref{metandsoft}, \kinms\ was given the intrinsic spatial and spectral resolution of MUSE, the pixel sampling of the MUSE cubes, and clouds were placed to best reproduce the respective moment 0 image of the emission line, in order to make the most realistic simulated cubes. Velocity fields of these simulated cubes were then created in python using the \textit{spectral-cube} package.
The residual (observed - simulated) velocity map was then used with several masking schemes and several metrics, to determine the simulated velocity map which best fits the observed velocity map.

\begin{figure}
\includegraphics[scale=0.35,clip,angle=0]{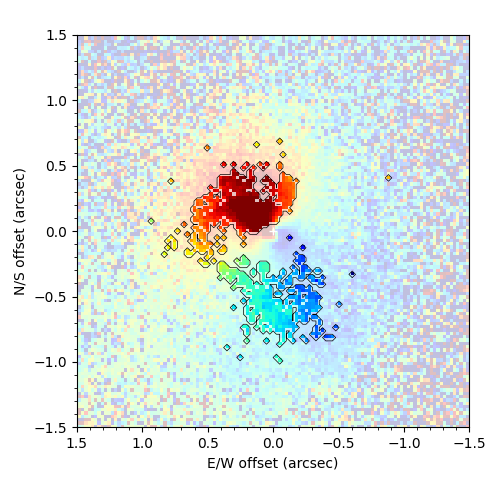}
\includegraphics[scale=0.34,clip,angle=0]{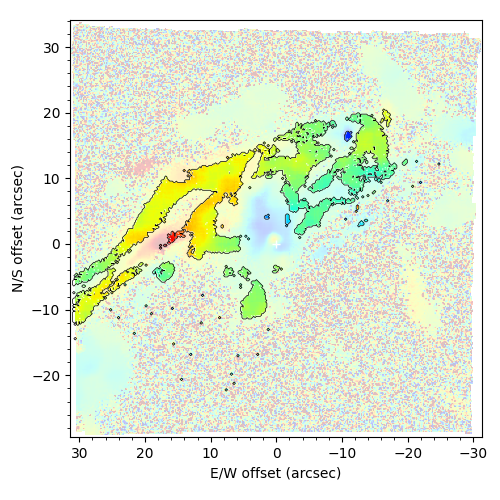}
\centering
\caption{Examples of masks applied to the subtracted velocity maps. The masks were made removing pixels with velocity dispersion larger than a given limit (sigma-clip mask). Left to right are the masks for \fulloi\ and \fullhanii. See Fig. \ref{fig:maskingapp} for other types of masks.}
\label{fig:masking}
\end{figure}

As emphasized in several previous sections, both the nuclear ionized disk and the large scale ionized gas emission show complex kinematics, often due to the superposition of several distinct components. The masking used on the residual velocity field before calculating best-fit metrics is thus important. We used several masks:
(a) Gauss-fit mask: all spaxels which show large residuals between the observed spectrum and the best fit Gaussian were masked.  This effectively masked both low SNR spaxels and spaxels with complex profiles which could not be fit with a single Gaussian;
(b) sigma-clip mask: all pixels in which the \oi\ dispersion map (Fig. \ref{fig:oimoms}) has values larger than 350 \kms\ were masked. This masks also effectively avoids the inclusion of spaxels with multiple velocity components;
(c) spiral-arm mask: all pixels which fall within the mini spiral arms seen in the \oi\ velocity residual map for the HBH i25 model (lower right panel of Fig. \ref{fig:oimoms}) were masked. Specifically we masked pixels with absolute residual velocities larger than 100 \kms;
(d) annular mask: all pixels outside the annular region between 0\farcs2 and 0\farcs4 from the nucleus, and pixels inside a rectangular region between $-$0\farcs6 and 0\farcs1 in E/W offset, and $-$0\farcs45 and 0\farcs15 in N/S offset, were masked. This mask retains the highest SNR regions of the disk while deleting the innermost region (with isophotal twists) and the high velocity dispersion nuclear region. 
The sigma-clip mask is shown in Fig. \ref{fig:masking}; the remaining masks can be found in Appendix \ref{fig:maskingapp}.

At each value of SMBH mass and inclination, we calculated several metrics from the masked velocity residual (observed - simulated) map; multiple metrics were used in order to be robust against changes in the histogram profile and average velocity in a given residual map. These included the standard deviation, the mean of the absolute difference, the difference between the  84 and 16 percentiles, and the r squared value (the last  constructed using the observed and model velocities rather than the distribution of the residual velocities).  These metrics were then evaluated across the parameter space of SMBH mass and inclination. Most combinations of masks and metrics deliver similar results with respect to favored pairs of SMBH mass and disk inclination, so we show only selected examples here.
The exercise above was also made for a comparison of the observed \oi\ velocity map with simulated cubes of the RIAF model of \citet[][see Sect. \ref{metandsoft}]{jet19}. 

Figure \ref{fig:mbhioi} presents the resulting metric obtained from the \oi\ emission line residual velocity maps for a Keplerian rotation model (left panels) and for the Jeter RIAF model (right panels). Specifically, here we show the results of the standard deviation of the residual velocity maps for three different masks: Gauss-fit, sigma-clip, and spiral-arm. 
We concentrate first on the left panels in Fig. \ref{fig:mbhioi}, i.e., 'Keplerian rotation' models. In general the minimum values of standard deviation follow a 'banana' pattern, with almost equal support for high black hole masses at low inclinations as for low black hole masses at higher inclinations, with black hole mass decreasing with increasing inclination. 
The uppermost panel is the result from the Gauss-fit mask. It favors a high black hole mass and small inclination, with a minimum beyond the range of the plot; the masking here does not necessarily exclude all high dispersion regions and spiral arms seen in the moment (and residual) maps.
The sigma-clip and spiral-arm masks (central and lower panels) yield similar results. Both plots show a banana-shaped dark blue region of minimum values, where the metric does not vary significantly. 
For our specific LBH and HBH values, the inclinations at minimum standard deviation are,  respectively, $\sim$37$\dg$ and $\sim$28$\dg$. 
There is very slight support for an HBH at lower inclination: 
the standard deviations (for the sigma-clip mask) for the LBH with i = 37$\dg$ and HBH with i = 28$\dg$ are 62.2 and 64.8 respectively, a difference of 4.2\%; for the spiral-arm mask, the difference is 3.92\%, in the same sense. 
In general, over most masks and metrics, a HBH at low inclination is favored over a LBH at high inclination only at the $\sim$4\% level. 

The equivalent results for the RIAF model residuals are shown in the right panels of Fig. \ref{fig:mbhioi}. As with the Keplerian model, we see a dark blue banana-shaped minimum region, but this is offset with respect to the Keplerian results as expected:  sub-Keplerian rotation results in higher inclinations favored for a given black hole mass.
Nevertheless, the level of sub-Keplerian rotation in this specific RIAF model is not sufficient to support the scenario of a HBH at inclination 42$\dg$:  values of $\Omega$ significantly lower than $\sqrt{0.7}$ would be required. 
Once more the Gauss-fit mask (uppermost panel) favors a high black hole mass and low inclination beyond the plot.
For the sigma-clip and spiral-arm masks the inclination ranges for a minimum standard deviation for the LBH and HBH are, respectively, $\sim$40$\dg$--43$\dg$ and $\sim$30$\dg$--31$\dg$. 
The respective standard deviations are 76.5 and 68.7 for the sigma-clip mask, and
56.4 and 54.4 for the spiral-arm mask.
The standard deviation for the HBH case is thus lower by 11\% for the sigma-clip mask and 4\% for the spiral-arm mask. 
Thus overall, within the Jeter RIAF model, a HBH in a low inclination disk provides a  slightly better fit than an LBH in a higher inclination disk, and a HBH in a disk with inclination i = 42$\dg$ is not supported. 

Intercomparing 'Keplerian disk' scenarios with the Jeter RIAF scenario, we note the following: 
(a) for a HBH,  the 'Keplerian disk' scenario gives lower velocity residuals than the RIAF model across all masks and metrics; 
(b) interestingly, a LBH in a i = 42$\dg$ disk is better supported in a RIAF scenario as compared to the 'Keplerian disk' scenario, and larger values of $\Omega$ would make the RIAF model even more favorable. 

\begin{figure}
\includegraphics[scale=0.35,clip,angle=0]{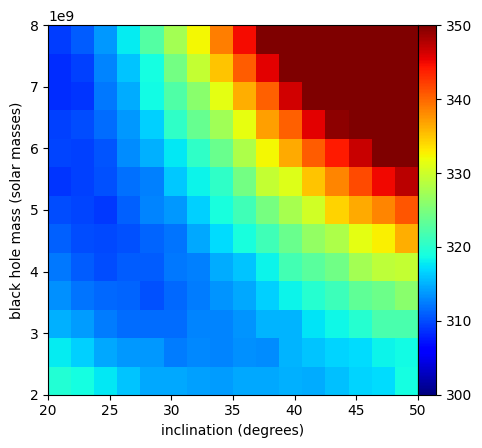}
\includegraphics[scale=0.35,clip,angle=0]{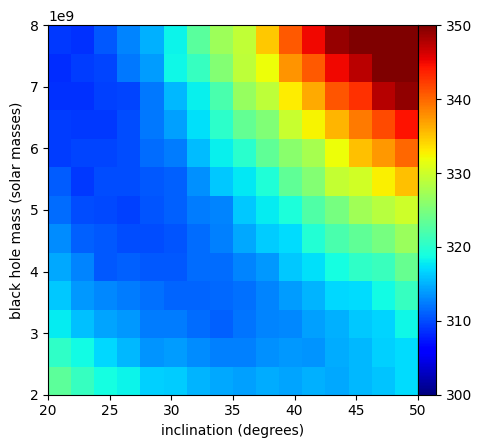}
\includegraphics[scale=0.35,clip,angle=0]{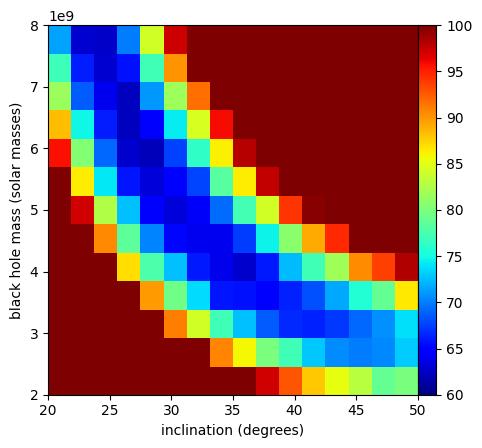}
\includegraphics[scale=0.35,clip,angle=0]{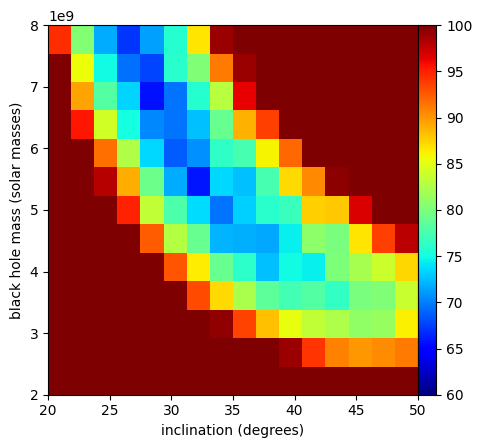}
\includegraphics[scale=0.35,clip,angle=0]{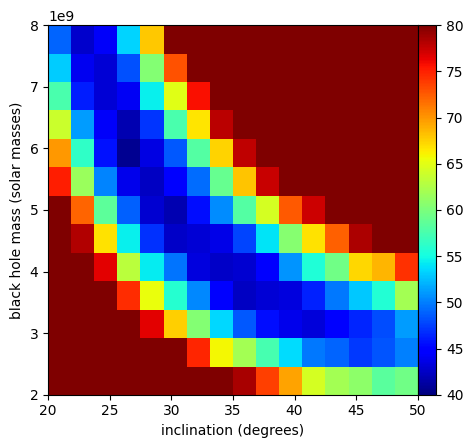}
\includegraphics[scale=0.35,clip,angle=0]{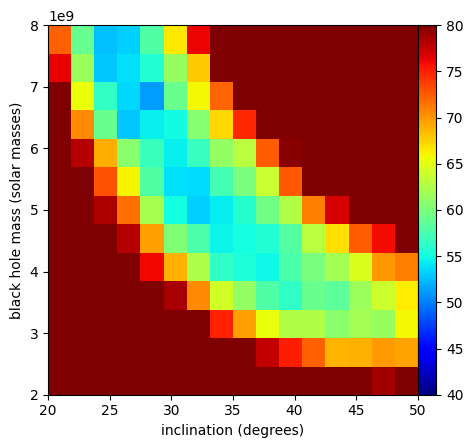}
\centering
\caption{Standard deviation (in \kms) of values in the residual (observed minus \kinms\ model) velocity map, as function of the black hole mass and disk inclination of the model, following the color bar to the right of each panel. The left column is for the 'Keplerian disk' model, and the right column for the Jeter RIAF model. Top to bottom are the standard deviations measured using Gauss-fit, sigma-clip and spiral-arm masks. The panels of each row follow the same color bar.}
\label{fig:mbhioi}
\end{figure}

\begin{figure}
\includegraphics[scale=0.35,clip,angle=0]{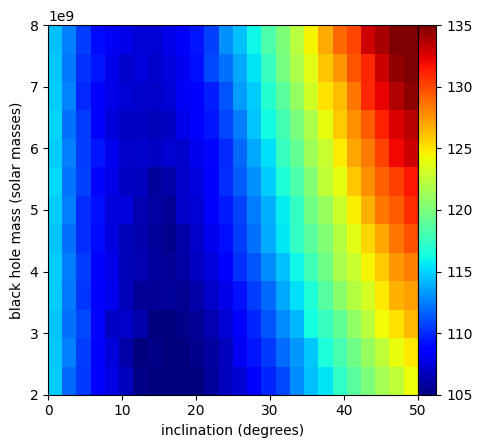}
\includegraphics[scale=0.35,clip,angle=0]{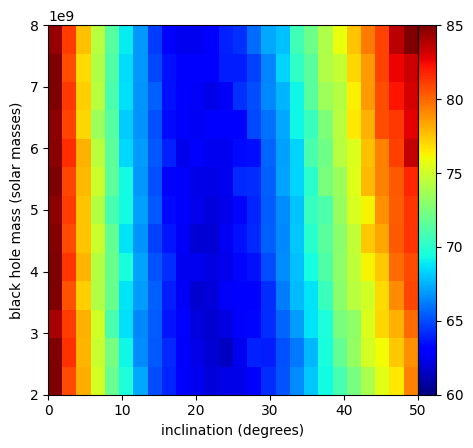}
\centering
\caption{As Fig. \ref{fig:mbhioi} but for the \wfm\ \fullhanii\ velocity map (i.e. the extended filaments) and the 'Keplerian' rotation model. Both panels show the standard deviation of the residual velocity map, with the left panel showing the result with the Gauss-fit mask and the right panel for the sigma-clip mask.}
\label{fig:mbhinii}
\end{figure}

We repeated the exercise for the velocity map of the large scale filaments seen in the \wfm\ cube, and the masking (a) and (b) explained for \oi. Here we used the velocity map of the \hanii\ line created via Gaussian fitting (with a single Gaussian fit to each of the \ha\ and \nii\ lines). 
The only differences from the process above is that (a) we varied the inclination between 0\arcdeg\ to 50\arcdeg; (b) the sigma-clip mask was created by masking pixels with velocity dispersion above 100 \kms; and (c) we only used the 'Keplerian' rotation models.
The resulting standard deviation in the residual velocity map, for intensity masking (left) and velocity dispersion masking (right) are shown in Fig. \ref{fig:mbhinii}. 
In both cases, the minima are seen at relatively small inclinations, almost independent of the black hole mass. This is since most of the non-masked area is outside of the sphere of influence of the SMBH. The minimum standard deviation is seen at an inclination of about 18\arcdeg\ for the Gauss-fit mask and 24\arcdeg\ for the sigma-clip mask; using the other metrics and masks give similar results. This supports the scenario that the filaments are rotating in a disk with inclination $\sim$20\arcdeg, close to the value we support for the inner sub-arcsec disk.

\section{Ionized gas outflow}
\label{igoutflow}
 
The moment maps and pv diagrams in the diffuse blue region and filaments to the N of the nucleus clearly demonstrate the presence of a non-rotating component. Additionally, some excessively redshifted gas is seen $\sim$3\arcsec\ SW of the nucleus. With projected velocities up to 400 \kms, the most viable explanation for these velocities is an AGN driven outflow.

The left panel of Fig. \ref{fig:cones} shows the residual velocity map (observed velocity minus our HBH i25 rotating disk model) of the \niitwo\ line, and emphasizes better not only the blue and redshifted regions to the N and S, respectively, but also the velocity gradients seen in these.

One could attempt to reproduce these velocity features with a (fully-filled) biconical outflow. Illustrative examples of these are shown in Fig. \ref{fig:appbicone}, where the cone axes were chosen to illustratively 'match' the blueshifted and redshifted regions, or to project onto the PA of the jet.
Neither of these illustrative models can satisfy the observed velocities while maintaining an origin in an AGN produced outflow along the axis of the jet. 

Instead we invoke a biconical outflow in which the cones are only partially filled with ionized gas; a scenario supported by the filamentary and patchy nature of the ionized gas in the inner 30\arcsec. To illustrate this scenario, Fig. \ref{fig:cones} shows the velocity structure of each cone individually. Here both cones have the same opening angle (45\arcdeg), and their orientation is chosen so that both cone axes project to PA $-$72\arcdeg\ (the PA of the jet). The velocity gradients in each bicone match those seen in the residual velocity map (left panel of Fig. \ref{fig:cones}). 
Thus, supported by the non-axisymmetric filamentary and patchy ionized gas seen on scales larger than 1\arcsec\ in both the \nfm\ and \wfm\ cubes, we posit that the diffuse and filamentary ionized gas to the N (outside the accretion disk and within the \nfm\ FOV) is primarily between the nucleus and the observer, and the ionized gas to the SW of the nucleus (outside the accretion disk and within the \nfm\ FOV) is behind the nucleus from the observers point of view. This partial filling of the approaching and receding cones then well recreates the velocity residual map of Fig. \ref{fig:cones}. 
This posited partially filled biconical outflow could also explain two other ionized gas blobs to the S and SE beyond the NFM FOV, but observed in the WFM FOV residual velocity map (bottom right panel of Fig. \ref{fig:wfmmoms}).

\begin{figure*}
\includegraphics[scale=0.53,clip,angle=0]{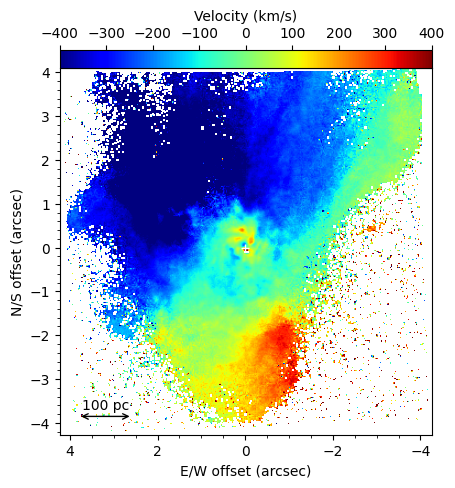}
\includegraphics[scale=0.53,clip,angle=0]{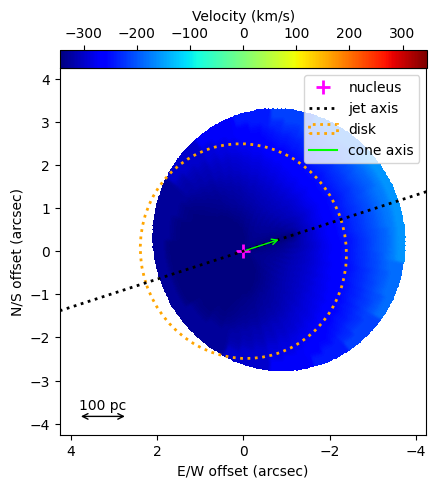}
\includegraphics[scale=0.53,clip,angle=0]{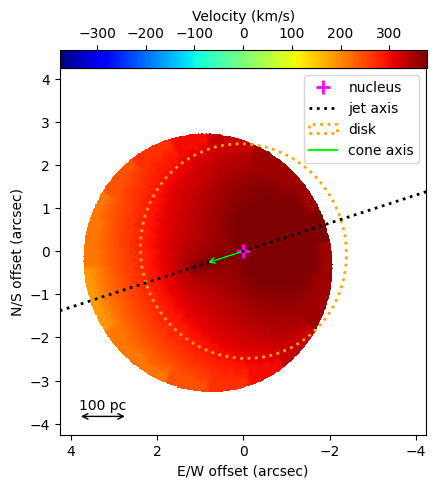}
\centering
\caption{Modeling of the ionized gas outflow in \gal.
Left panel: residual (after subtracting the HBH i25 model for the nuclear disk) \nii\ velocity map. We emphasize the gradients in the blue and redshifted velocities to the NE and S. 
Middle and right panels: models of a filled conical outflow made with \kinms, with each cone showed separately. The cone half-opening angles are 45\arcdeg\ (approaching; middle panel) and 30\arcdeg\ (receding; right panel). Each cone has an extent of 3\arcsec\ along its axis, and the radial outflow velocity within each cone is 400 \kms. The inclination of both cone axes to the line of sight is 18\arcdeg. The green line shows the projection of the cone axes on the plane of the sky; these project to PA 288\arcdeg, the PA of the jet axis. The orange dotted lines delineate the projected shape of a disk perpendicular the cone axes.
}
\label{fig:cones}
\end{figure*}

\section{Discussion and Conclusions}
\label{discussion}

Our full areal coverage of the nuclear (new deep MUSE-AO \nfm\ cube) and larger scale (archival MUSE \wfm\ cube) ionized gas, allow improved constraints on the morphology and kinematics of the ionized gas in the inner 0\farcs1 to 30\arcsec\ in \gal. 
On the larger scales, filaments W1 and W2 (which are 10s of arcsec in extent) show velocities which can, in most of their extent, be fit with a partially filled gas disk in  inclination i $\sim$25$\dg$ rotating in the potential of the galaxy. Nevertheless, certain sections of these filaments show large kinks and highly dispersive velocities. The W arm of W1 appears to cross the nucleus in projection, and at relatively low rotation-corrected velocity. While it cannot be clearly traced between 1\arcsec--2\arcsec\ W of the nucleus, its incoming direction aligns with a high velocity residual (after rotation correction) region in the SW quadrant of the arcsec-scale ionized disk. We speculate that this may be the point at which the filament W1 feeds the disk.

We posit the presence of a wide angle (half-opening angle of $\sim$45\arcdeg) outflow along the same (3 dimensional) axis as the jet.
The diffuse and blue filamentary structures on scales of \ias{1.5}{4} to the N participate in this conical radial outflow (and are thus in front of the nucleus). Equivalently, the redshifted gas at the same distance to the SW is behind the nucleus and part of the redshifted conical outflow. 
Apart from this velocity structure, the conical outflow model is supported by the velocity gradients seen in the blueshifted (N) and redshifted (S) ionized gas seen in the regions \ias{1}{4} from the nucleus. With a radial velocity of $\sim$400 \kms, the outflow requires to be powered by the AGN, but since the blue and red shifted cones are only partially filled with ionized gas, the ionized gas would have to be entrained from the patchy and filamentary ISM by a wide angle wind launched near the black hole. Each cone requires to be at least 3\arcsec\ in extent, and thus the crossing time is 0.6 Myr, significantly shorter than the lifetime of the \gal\ larger scale jet and lobes \citep{Owen2000}. 

Many of the filaments seen in the inner 4\arcsec\ appear to cross the nucleus. We see all of the following three cases (a) those which follow the potential of the black hole, and thus likely truly enter the inner 5 pc; (b) those which enter the nucleus, but only in projection, as their bulk blueshifts are not changed as (projected) nuclear separation decreases; and (c) at least one mixed case where highly blueshifted filament participates in a systemic-velocity-offset quasi-Keplerian rotation as it crosses the nucleus in projection; for illustration we modeled this as Keplerian rotation in a plane with inclination = 60$\dg$. Two potential explanations are a gas filament with a large bulk velocity which approaches the nucleus at high inclination, or Keplerian rotating gas near the nucleus which is swept into the biconical outflow.

With respect to the inner ionized gas disk: if the \gal\ jet is intrinsically two-sided, then the (brighter due to Doppler boosting) NW jet is oriented toward us. For the inner ($<$0.1\arcsec) accretion disk to be perpendicular to the jet axis, we expect that the SE half of the disk is the near side, and the NW the far side. If this is maintained, i.e. if the disk warp is 'slight', we expect the same for the arcsecond scale ionized gas disk. 
Indeed, dust lanes seen in the HST images of \citet{for94} led them to suggest that the SE side is the near side. 

The twisted isophotes seen in the inner arcsec of the ionized gas disk potentially resolve another long standing problem: the mismatch between the axes of the inner ionized gas disk and the prominent jet. This issue is extensively discussed in \citet{jet19}. Our velocity maps in Fig. \ref{fig:oimoms} provide clear evidence for twisted velocity isophotes in the inner arcsecond of the disk. The twist is  clearest from PA $\sim$42$\dg$ at 0\farcs3--0\farcs5\ to PA $\sim$90$\dg$ at 0\farcs1--0\farcs2. Inside this radius, there is a suggestion of a further twist: the value of this innermost PA is most clearly constrained using the residual maps of our best fit model (bottom right panel of Fig. \ref{fig:oimoms}). Here, the residual over-subtraction is in PA $\sim$17\arcdeg,  perpendicular to the jet axis. Since our innermost resolved region is about $\sim$5 pc, there is of course room for further twists down to the jet launching region around the black hole. For radii $\geq$0\farcs3, the redshifted side of the disk shows some indications for further twists in the velocity isophotes, though at 10s of arcsecond scales we again support a PA close to 40$\dg$ and inclination equal to the inner arcsec disk. 

Our kinematic analysis clearly shows the presence of multiple velocity components (disk, outflow, and filaments) overlapping in the nucleus. Given this, it is dangerous to constrain the inclination of the disk using isophotal fits. Our \hanii\ and \oi\ moment 0 maps do not show a clear and consistent inclined disk shape. In these, while the different intensity levels are highly non axi-symmetric, one gets more of an impression of a face-on, rather than inclined, disk. 
At least a part of this face-on appearance could be due to filament 8 passing through the nucleus in projection. Its 'zero' velocity gas would extend the intensity contours of the disk along its minor axis. To the N ionized gas in the outflow is seen even in the inner arcsec, elongating the N-S extent at some radii. 
Morphologically, the disk is best constrained via its spiral arm structure, which we see both in intensity and kinematics. These spiral arms are better encompassed by a i = 25\arcdeg, rather than i = 42\arcdeg\ disk. The kinematics of the \oi\ line in the  inner 0\farcs2 to 0\farcs6 is slightly better fit with an inclination of $\sim$25\arcdeg\ as compared to 42\arcdeg. Finally, we have also shown that the kinematics of the large scale filaments W1 and W2 which extend from a few to tens of arcseconds from the nucleus, can be reasonably explained in most parts by a rotating disk with i = 25\arcdeg\ (but not as well or consistently with a i = 42\arcdeg\ disk). Together, the three results favor a relatively face-on disk rather than one at i $\sim$42\arcdeg. This lower inclination disk is also consistent with the posited 3-dimensional direction of the jet. Though, given 
the warp in the PA of the disk, it is impossible to rule out a change in inclination inside our central resolution elements. 

A precise measurement (with well defined errors) of the black hole mass via ionized gas kinematics is not possible due to the many morphological and kinematic complexities in the ionized gas, which  have been discussed extensively in this work. Even removing the effects of the ionized filaments which cross the nucleus (sometimes only in projection), and the outflow component, the irregular shape of the disk, the non axis-symmetric velocities displayed by the disk, the twisted inner velocity isophotes, and the finding that the spiral arms are slightly redshifted from the diffuse gas in the inner disk, all preclude a sufficiently accurate determination of the inclination of the subarcsec-scale disk, and thus black hole mass. 

Nevertheless, the deep and integral spectroscopy from MUSE allow us to clarify and constrain previous contradictory black hole mass measurements: e.g., the high mass black holes of G11, the EHT Collaboration, and \citet{lie23} versus the low mass black hole of W13. First and foremost, we have shown that the complexities of the nuclear ionized gas highly complicates the determination an accurate measurement of the black hole mass from ionized gas kinematics, which immediately lends more weight to the high mass value. However, some indications can be leveraged from the new data, especially via the \oi\ velocities. With the caveat of all the complexities involved,
we find no reason to favor a i = 42$\dg$ disk over a lower (i $\sim$25$\dg$) one. However, several factors support, even if sometimes weakly, a disk inclination closer to 25\arcdeg:
(a) fits to the sub-arcsec ionized gas disk with the \kinemetry\ package supports inclinations closer to 20$\dg$--25$\dg$, a PA of $\sim$45$\dg$, with the disk velocities dominated by rotation; 
(b) a minimization of the differences between the observed velocity fields of the sub-arcsec ionized gas disk and simulated velocity fields derived via \kinms\ over a range of black hole mass and inclination, lends weak support to the scenario of a high mass black hole in a low inclination disk, as opposed to a low mass black hole in a higher (42$\dg$) inclination disk;
(c) visual comparisons of  the LBH i42 model and the HBH i25 model with the observed velocity fields and pv diagrams, also weakly support the HBH i25 model, though their predictions are very similar within the complexities, and large linewidths, of the emission lines;
(d) morphologically, the spiral arms in the inner arcsec of the ionized disk, which are the most plausible parts of the 'disk' involved in circular rotation, are better encompassed by a i = 25\arcdeg, rather than i = 42\arcdeg\ disk; 
(e) the outer ionized gas filaments on 10s of arcsec scales can be relatively well fit with a rotating disk with i = 25\arcdeg, but not with i = 42\arcdeg. At these large nuclear distances, the rotation is dominated by the galaxy potential, so that there is no longer a degeneracy between black hole mass and inclination in our rotational fits. 
Finally, and independently, we note that the lower inclination disk provides a more consistent alignment between the axes of the ionized gas disk, the well known jet, and the biconical outflow we find here. 
Thus the most consistent explanation for the observed inner ionized gas rotating disk is a black hole with mass closer to 6.0 $\times$ 10$^9$ \msun\ surrounded by a low (i $\sim$25$\dg$) inclination disk. 

To reduce the tension between the G11 and W13 mass measurements, \citet{jet19} have suggested intrinsic sub-Keplerian disk rotation of the ionized gas as a potential solution. We tested the specific RIAF model proposed by these authors (over a range of black hole masses and disk inclinations) against the observed velocity field of the sub-arcsec nuclear disk, and found the following: 
(a) the sub-Keplerian rotational velocities of RIAF models indeed allow a higher black hole mass in a given disk inclination as compared to Keplerian rotation models, but the specific value of $\Omega$ in the Jeter RIAF model is not sufficiently small to allow a HBH in a i = 42$\dg$ RIAF disk to masquerade as a LBH in a i = 42$\dg$ Keplerian disk. 
(b) the residual (observed minus model) velocity fields of Keplerian models were compared to those with RIAF  models. Except for one masking scheme used by us (Gauss-fit masking), Keplerian models result in significantly smaller velocity residuals as compared to  RIAF models when both black hole mass and inclination can freely vary. 
Reducing the value of $\Omega$ can solve the issues in item (a) above, but not item (b). Further, the disk outflow in the  RIAF model should result in significant non-circular motions in the disk. However our \kinemetry\ fits to the sub-arcsec ionized gas disk implies that the bulk of the velocities can be fit with pure rotation, i.e.  non-rotative contributions are small. Thus overall, a HBH in a low inclination disk is a simpler and better-fit alternative to a HBH in a RIAF inflow.

Putting our results together gives us a simple and consistent picture of the nucleus of \gal. A black hole mass closer to $\sim$6.0--6.5 (rather than 3.5) billion solar masses (for a distance of 16.8 Mpc), together with a warped (in PA at least) sub-arcsec ionized disk with inclination $\sim$25\arcdeg. The PA of the innermost accretion disk seen in our residual \oi\ velocity map suggests a disk axis aligned with the axes of both the jet and the (partially filled) nuclear radial biconical outflow. The disk inclination to the line of sight matches that expected from Doppler boosting of the approaching jet. The kinematics of the ionized filaments at 4\arcsec--20\arcsec\ are reasonably well explained by a (partially filled) rotating disk with inclination 25\arcdeg, though the several kinks and bubbles along their length point to the presence of other flow velocities and instabilities. 

The MUSE \wfm\ and \nfm\ cubes are synergetically useful as they can also be used to extract the stellar dynamics on scales of 0\farcs1 to 30\arcsec. \sch\ modeling of the observed intensity, velocity, and line of sight velocity distribution of the stars is deferred to Osorno et al., in preparation. Further, the MUSE cubes, together with multi-epoch and multi-filter HST imaging, provide a deep insight into the (patchy) dust structures, the ionized gas distribution, and their relationship with the pc-scale to kpc-scale radio jet. These are being explored in Richtler et al., in preparation. 

\begin{acknowledgements}
This work was funded by the National Agency for Research and Development (ANID). 
JO acknowledges support from / Scholarship Program / Doctorado Nacional / 2017-21171690. NN and JO acknowledge funding from ANID via Nucleo Milenio TITANs (NCN19$-$058), Fondecyt 1221421, and Basal FB210003.
PH is a member of, and received financial support from, the International Max Planck Research School (IMPRS) for Astronomy and Astrophysics at the Universities of Bonn and Cologne.
Based on observations collected at the European Southern Observatory under ESO programme 0103.B-0581.

\end{acknowledgements}

%
%

\bibliographystyle{aa}
\bibliography{mainreferences}

\begin{thebibliography}{47}
\expandafter\ifx\csname natexlab\endcsname\relax\def\natexlab#1{#1}\fi

\bibitem[{{Bacon} {et~al.}(2010){Bacon}, {Accardo}, {Adjali}, {Anwand},
  {Bauer}, {Biswas}, {Blaizot}, {Boudon}, {Brau-Nogue}, {Brinchmann},
  {Caillier}, {Capoani}, {Carollo}, {Contini}, {Couderc}, {Daguis{\'e}},
  {Deiries}, {Delabre}, {Dreizler}, {Dubois}, {Dupieux}, {Dupuy}, {Emsellem},
  {Fechner}, {Fleischmann}, {Fran{\c{c}}ois}, {Gallou}, {Gharsa}, {Glindemann},
  {Gojak}, {Guiderdoni}, {Hansali}, {Hahn}, {Jarno}, {Kelz}, {Koehler},
  {Kosmalski}, {Laurent}, {Le Floch}, {Lilly}, {Lizon}, {Loupias}, {Manescau},
  {Monstein}, {Nicklas}, {Olaya}, {Pares}, {Pasquini}, {P{\'e}contal-Rousset},
  {Pell{\'o}}, {Petit}, {Popow}, {Reiss}, {Remillieux}, {Renault}, {Roth},
  {Rupprecht}, {Serre}, {Schaye}, {Soucail}, {Steinmetz}, {Streicher}, {Stuik},
  {Valentin}, {Vernet}, {Weilbacher}, {Wisotzki}, \& {Yerle}}]{muse}
{Bacon}, R., {Accardo}, M., {Adjali}, L., {et~al.} 2010, 7735, 773508

\bibitem[{{Barth} {et~al.}(2016){Barth}, {Darling}, {Baker}, {Boizelle},
  {Buote}, {Ho}, \& {Walsh}}]{bar16}
{Barth}, A.~J., {Darling}, J., {Baker}, A.~J., {et~al.} 2016, \apj, 823, 51

\bibitem[{{Bentz} {et~al.}(2010){Bentz}, {Walsh}, {Barth}, {Yoshii}, {Woo},
  {Wang}, {Treu}, {Thornton}, {Street}, {Steele}, {Silverman}, {Serduke},
  {Sakata}, {Minezaki}, {Malkan}, {Li}, {Lee}, {Hiner}, {Hidas}, {Greene},
  {Gates}, {Ganeshalingam}, {Filippenko}, {Canalizo}, {Bennert}, \&
  {Baliber}}]{ben10}
{Bentz}, M.~C., {Walsh}, J.~L., {Barth}, A.~J., {et~al.} 2010, \apj, 716, 993

\bibitem[{{Bittner} {et~al.}(2019){Bittner}, {Falc{\'o}n-Barroso}, {Nedelchev},
  {Dorta}, {Gadotti}, {Sarzi}, {Molaeinezhad}, {Iodice}, {Rosado-Belza}, {de
  Lorenzo-C{\'a}ceres}, {Fragkoudi}, {Gal{\'a}n-de Anta}, {Husemann},
  {M{\'e}ndez-Abreu}, {Neumann}, {Pinna}, {Querejeta},
  {S{\'a}nchez-Bl{\'a}zquez}, \& {Seidel}}]{gist}
{Bittner}, A., {Falc{\'o}n-Barroso}, J., {Nedelchev}, B., {et~al.} 2019, \aap,
  628, A117

\bibitem[{{Boselli} {et~al.}(2019){Boselli}, {Fossati}, {Longobardi},
  {Consolandi}, {Amram}, {Sun}, {Andreani}, {Boquien}, {Braine}, {Combes},
  {C{\^o}t{\'e}}, {Cuillandre}, {Duc}, {Emsellem}, {Ferrarese}, {Gavazzi},
  {Gwyn}, {Hensler}, {Peng}, {Plana}, {Roediger}, {Sanchez-Janssen}, {Sarzi},
  {Serra}, \& {Trinchieri}}]{bos19}
{Boselli}, A., {Fossati}, M., {Longobardi}, A., {et~al.} 2019, \aap, 623, A52

\bibitem[{{Cappellari}(2017)}]{ppxf2}
{Cappellari}, M. 2017, \mnras, 466, 798

\bibitem[{{Cappellari} \& {Copin}(2003)}]{vorbin}
{Cappellari}, M. \& {Copin}, Y. 2003, \mnras, 342, 345

\bibitem[{{Cappellari} \& {Emsellem}(2004)}]{ppxf1}
{Cappellari}, M. \& {Emsellem}, E. 2004, \pasp, 116, 138

\bibitem[{{Davis} {et~al.}(2013){Davis}, {Alatalo}, {Bureau}, {Cappellari},
  {Scott}, {Young}, {Blitz}, {Crocker}, {Bayet}, {Bois}, {Bournaud}, {Davies},
  {de Zeeuw}, {Duc}, {Emsellem}, {Khochfar}, {Krajnovi{\'c}}, {Kuntschner},
  {Lablanche}, {McDermid}, {Morganti}, {Naab}, {Oosterloo}, {Sarzi}, {Serra},
  \& {Weijmans}}]{kinms}
{Davis}, T.~A., {Alatalo}, K., {Bureau}, M., {et~al.} 2013, \mnras, 429, 534

\bibitem[{{Di Matteo} {et~al.}(2005){Di Matteo}, {Springel}, \&
  {Hernquist}}]{dim05}
{Di Matteo}, T., {Springel}, V., \& {Hernquist}, L. 2005, \nat, 433, 604

\bibitem[{{Emsellem} {et~al.}(2004){Emsellem}, {Cappellari}, {Peletier},
  {McDermid}, {Bacon}, {Bureau}, {Copin}, {Davies}, {Krajnovi{\'c}},
  {Kuntschner}, {Miller}, \& {de Zeeuw}}]{ems04}
{Emsellem}, E., {Cappellari}, M., {Peletier}, R.~F., {et~al.} 2004, \mnras,
  352, 721

\bibitem[{{Emsellem} {et~al.}(2014){Emsellem}, {Krajnovic}, \& {Sarzi}}]{ems14}
{Emsellem}, E., {Krajnovic}, D., \& {Sarzi}, M. 2014, \mnras, 445, L79

\bibitem[{{Event Horizon Telescope Collaboration}(2019)}]{ehtc19}
{Event Horizon Telescope Collaboration}. 2019, \apjl, 875, L6

\bibitem[{{Event Horizon Telescope Collaboration}(2022)}]{ehtc21}
{Event Horizon Telescope Collaboration}. 2022, \apjl, 930, L15

\bibitem[{{Falc{\'o}n-Barroso} {et~al.}(2006){Falc{\'o}n-Barroso}, {Bacon},
  {Bureau}, {Cappellari}, {Davies}, {de Zeeuw}, {Emsellem}, {Fathi},
  {Krajnovi{\'c}}, {Kuntschner}, {McDermid}, {Peletier}, \& {Sarzi}}]{gandalf2}
{Falc{\'o}n-Barroso}, J., {Bacon}, R., {Bureau}, M., {et~al.} 2006, \mnras,
  369, 529

\bibitem[{{Ferrarese} \& {Merritt}(2000)}]{fer00}
{Ferrarese}, L. \& {Merritt}, D. 2000, \apjl, 539, L9

\bibitem[{{Ford} {et~al.}(1994){Ford}, {Harms}, {Tsvetanov}, {Hartig},
  {Dressel}, {Kriss}, {Bohlin}, {Davidsen}, {Margon}, \& {Kochhar}}]{for94}
{Ford}, H.~C., {Harms}, R.~J., {Tsvetanov}, Z.~I., {et~al.} 1994, \apjl, 435,
  L27

\bibitem[{{Gao} {et~al.}(2017){Gao}, {Braatz}, {Reid}, {Condon}, {Greene},
  {Henkel}, {Impellizzeri}, {Lo}, {Kuo}, {Pesce}, {Wagner}, \& {Zhao}}]{gao17}
{Gao}, F., {Braatz}, J.~A., {Reid}, M.~J., {et~al.} 2017, \apj, 834, 52

\bibitem[{{Gavazzi} {et~al.}(2000){Gavazzi}, {Boselli}, {V{\'\i}lchez},
  {Iglesias-Paramo}, \& {Bonfanti}}]{gav00}
{Gavazzi}, G., {Boselli}, A., {V{\'\i}lchez}, J.~M., {Iglesias-Paramo}, J., \&
  {Bonfanti}, C. 2000, \aap, 361, 1

\bibitem[{{Gebhardt} {et~al.}(2011){Gebhardt}, {Adams}, {Richstone}, {Lauer},
  {Gultekin}, {Murphy}, {Faber}, \& {Tremaine}}]{geb11}
{Gebhardt}, K., {Adams}, J., {Richstone}, D., {et~al.} 2011, 217, 422.05

\bibitem[{{Gebhardt} \& {Thomas}(2009)}]{geb09}
{Gebhardt}, K. \& {Thomas}, J. 2009, \apj, 700, 1690

\bibitem[{{Gralla} {et~al.}(2020){Gralla}, {Lupsasca}, \&
  {Marrone}}]{Gralla2020}
{Gralla}, S.~E., {Lupsasca}, A., \& {Marrone}, D.~P. 2020, \prd, 102, 124004

\bibitem[{{G{\"u}ltekin} {et~al.}(2009){G{\"u}ltekin}, {Richstone}, {Gebhardt},
  {Lauer}, {Tremaine}, {Aller}, {Bender}, {Dressler}, {Faber}, {Filippenko},
  {Green}, {Ho}, {Kormendy}, {Magorrian}, {Pinkney}, \& {Siopis}}]{gul09}
{G{\"u}ltekin}, K., {Richstone}, D.~O., {Gebhardt}, K., {et~al.} 2009, \apj,
  698, 198

\bibitem[{{Harms} {et~al.}(1994){Harms}, {Ford}, {Tsvetanov}, {Hartig},
  {Dressel}, {Kriss}, {Bohlin}, {Davidsen}, {Margon}, \& {Kochhar}}]{har94}
{Harms}, R.~J., {Ford}, H.~C., {Tsvetanov}, Z.~I., {et~al.} 1994, \apjl, 435,
  L35

\bibitem[{{Jeter} {et~al.}(2019){Jeter}, {Broderick}, \& {McNamara}}]{jet19}
{Jeter}, B., {Broderick}, A.~E., \& {McNamara}, B.~R. 2019, \apj, 882, 82

\bibitem[{{Johannsen} \& {Psaltis}(2010)}]{Johannsen2010}
{Johannsen}, T. \& {Psaltis}, D. 2010, \apj, 718, 446

\bibitem[{{Kormendy} \& {Gebhardt}(2001)}]{kor01}
{Kormendy}, J. \& {Gebhardt}, K. 2001, in American Institute of Physics
  Conference Series, Vol. 586, 20th Texas Symposium on relativistic
  astrophysics, ed. J.~C. {Wheeler} \& H.~{Martel}, 363--381

\bibitem[{{Kormendy} \& {Ho}(2013)}]{kor13}
{Kormendy}, J. \& {Ho}, L.~C. 2013, \araa, 51, 511

\bibitem[{{Krajnovi{\'c}} {et~al.}(2006){Krajnovi{\'c}}, {Cappellari}, {de
  Zeeuw}, \& {Copin}}]{kinemetry}
{Krajnovi{\'c}}, D., {Cappellari}, M., {de Zeeuw}, P.~T., \& {Copin}, Y. 2006,
  \mnras, 366, 787

\bibitem[{Liepold {et~al.}(2023)Liepold, Ma, \& Walsh}]{lie23}
Liepold, E.~R., Ma, C.-P., \& Walsh, J.~L. 2023, Keck Integral-Field
  Spectroscopy of M87 Reveals an Intrinsically Triaxial Galaxy and a Revised
  Black Hole Mass

\bibitem[{{Macchetto} {et~al.}(1997){Macchetto}, {Marconi}, {Axon}, {Capetti},
  {Sparks}, \& {Crane}}]{mac97}
{Macchetto}, F., {Marconi}, A., {Axon}, D.~J., {et~al.} 1997, \apj, 489, 579

\bibitem[{{Marconi} \& {Hunt}(2003)}]{mar03}
{Marconi}, A. \& {Hunt}, L.~K. 2003, \apjl, 589, L21

\bibitem[{{Murphy} {et~al.}(2011){Murphy}, {Gebhardt}, \& {Adams}}]{mur11}
{Murphy}, J.~D., {Gebhardt}, K., \& {Adams}, J.~J. 2011, \apj, 729, 129

\bibitem[{{Owen} {et~al.}(2000){Owen}, {Eilek}, \& {Kassim}}]{Owen2000}
{Owen}, F.~N., {Eilek}, J.~A., \& {Kassim}, N.~E. 2000, \apj, 543, 611

\bibitem[{{Rusli} {et~al.}(2011){Rusli}, {Thomas}, {Erwin}, {Saglia}, {Nowak},
  \& {Bender}}]{rus11}
{Rusli}, S.~P., {Thomas}, J., {Erwin}, P., {et~al.} 2011, \mnras, 410, 1223

\bibitem[{{Saglia} {et~al.}(2016){Saglia}, {Opitsch}, {Erwin}, {Thomas},
  {Beifiori}, {Fabricius}, {Mazzalay}, {Nowak}, {Rusli}, \& {Bender}}]{sag16}
{Saglia}, R.~P., {Opitsch}, M., {Erwin}, P., {et~al.} 2016, \apj, 818, 47

\bibitem[{{Sargent} {et~al.}(1978){Sargent}, {Young}, {Boksenberg},
  {Shortridge}, {Lynds}, \& {Hartwick}}]{sar78}
{Sargent}, W.~L.~W., {Young}, P.~J., {Boksenberg}, A., {et~al.} 1978, \apj,
  221, 731

\bibitem[{{Sarzi} {et~al.}(2006){Sarzi}, {Falc{\'o}n-Barroso}, {Davies},
  {Bacon}, {Bureau}, {Cappellari}, {de Zeeuw}, {Emsellem}, {Fathi},
  {Krajnovi{\'c}}, {Kuntschner}, {McDermid}, \& {Peletier}}]{gandalf1}
{Sarzi}, M., {Falc{\'o}n-Barroso}, J., {Davies}, R.~L., {et~al.} 2006, \mnras,
  366, 1151

\bibitem[{{Sarzi} {et~al.}(2018){Sarzi}, {Spiniello}, {La Barbera},
  {Krajnovi{\'c}}, \& {van den Bosch}}]{sar18}
{Sarzi}, M., {Spiniello}, C., {La Barbera}, F., {Krajnovi{\'c}}, D., \& {van
  den Bosch}, R. 2018, \mnras, 478, 4084

\bibitem[{{Schnorr M{\"u}ller} {et~al.}(2011){Schnorr M{\"u}ller},
  {Storchi-Bergmann}, {Riffel}, {Ferrari}, {Steiner}, {Axon}, \&
  {Robinson}}]{mul11}
{Schnorr M{\"u}ller}, A., {Storchi-Bergmann}, T., {Riffel}, R.~A., {et~al.}
  2011, \mnras, 413, 149

\bibitem[{{Smith} {et~al.}(2021){Smith}, {Bureau}, {Davis}, {Cappellari},
  {Liu}, {Onishi}, {Iguchi}, {North}, {Sarzi}, \& {Williams}}]{wisdomvii}
{Smith}, M.~D., {Bureau}, M., {Davis}, T.~A., {et~al.} 2021, \mnras, 503, 5984

\bibitem[{{Vazdekis} {et~al.}(2010){Vazdekis}, {S{\'a}nchez-Bl{\'a}zquez},
  {Falc{\'o}n-Barroso}, {Cenarro}, {Beasley}, {Cardiel}, {Gorgas}, \&
  {Peletier}}]{miles}
{Vazdekis}, A., {S{\'a}nchez-Bl{\'a}zquez}, P., {Falc{\'o}n-Barroso}, J.,
  {et~al.} 2010, \mnras, 404, 1639

\bibitem[{{Walker} {et~al.}(2018){Walker}, {Hardee}, {Davies}, {Ly}, \&
  {Junor}}]{wal18}
{Walker}, R.~C., {Hardee}, P.~E., {Davies}, F.~B., {Ly}, C., \& {Junor}, W.
  2018, \apj, 855, 128

\bibitem[{{Walsh} {et~al.}(2013){Walsh}, {Barth}, {Ho}, \& {Sarzi}}]{wal13}
{Walsh}, J.~L., {Barth}, A.~J., {Ho}, L.~C., \& {Sarzi}, M. 2013, \apj, 770, 86

\bibitem[{{Werner} {et~al.}(2010){Werner}, {Simionescu}, {Million}, {Allen},
  {Nulsen}, {von der Linden}, {Hansen}, {B{\"o}hringer}, {Churazov}, {Fabian},
  {Forman}, {Jones}, {Sanders}, \& {Taylor}}]{wer10}
{Werner}, N., {Simionescu}, A., {Million}, E.~T., {et~al.} 2010, \mnras, 407,
  2063

\bibitem[{{Young} {et~al.}(1978){Young}, {Westphal}, {Kristian}, {Wilson}, \&
  {Landauer}}]{you78}
{Young}, P.~J., {Westphal}, J.~A., {Kristian}, J., {Wilson}, C.~P., \&
  {Landauer}, F.~P. 1978, \apj, 221, 721

\bibitem[{{Zhao} {et~al.}(2009){Zhao}, {Morris}, {Goss}, \& {An}}]{Zhao2009}
{Zhao}, J.-H., {Morris}, M.~R., {Goss}, W.~M., \& {An}, T. 2009, \apj, 699, 186

\end{thebibliography}

\begin{appendix}

\section{VLT/MUSE-NFM data versus HST/STIS data}
\label{app:stis-muse}

Previous ionized gas based measurements of the black hole mass in \gal\ used aperture \citep{har94} or slit spectra \citep{mac97,wal13} from instruments on HST.  In this work, we use MUSE IFU data cubes; the new MUSE data are superior for several reasons. MUSE is a relatively high throughput instrument on a 8 meter class telescope: in collecting area the VLT is factor $\sim$11 larger than HST. In adaptive optics mode, MUSE delivers a higher spatial resolution than STIS.
The STIS data used in \citet{wal13} consist of spectra in five parallel slits, one of them crossing the nucleus of the galaxy, whereas the MUSE \nfm\ datacube fully samples the inner \fov{8}{8}. The larger FOV and full spatial sampling allows a more comprehensive analysis along any PA.
The MUSE data also covers a larger wavelength range.
The only drawback of the dataset is that MUSE has a slightly lower spectral resolution as compared to previous HST instruments used for such studies.
However, in a massive elliptical like \gal, with relatively large velocities and velocity dispersions, the MUSE spectral resolution is sufficient to accurately resolve and measure gradients in these. 

Figure \ref{fig:stismuse} compares the pv diagrams obtained from the STIS data of \citet{wal13} and our MUSE data along a nuclear slit in PA 51\arcdeg\ (the slit PA used by W13). There is a improvement of the quality of the data, especially in the weaker \oione\ line.

\begin{figure}[h]
\centering
\includegraphics[width=8.15cm]{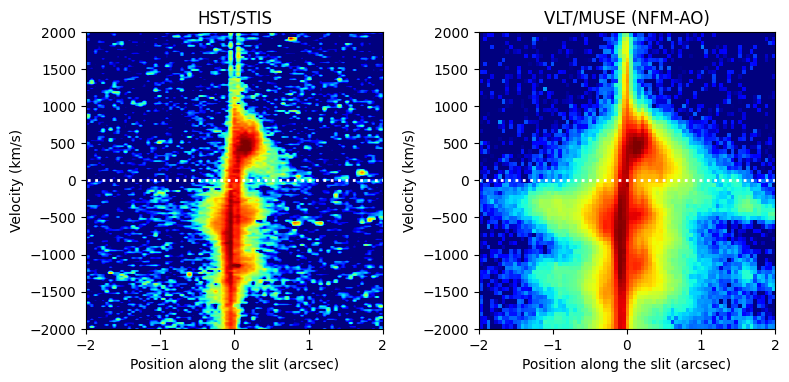} 
\includegraphics[width=8.15cm]{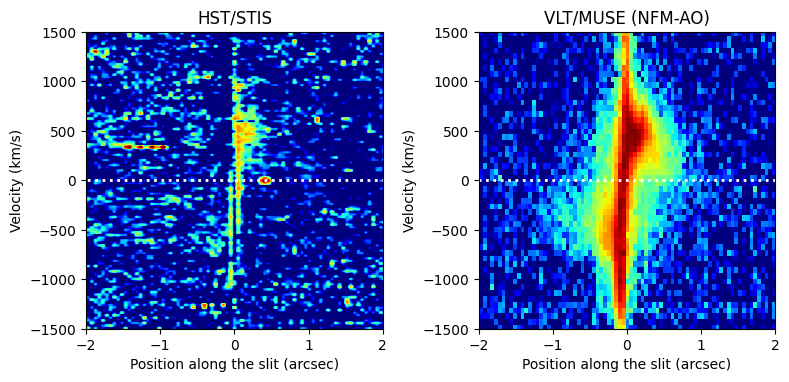} 
\caption{A comparison of the pv diagrams obtained from previous HST/STIS long-slit spectra and our VLT/MUSE-NFM-AO datacube for the \niitwo\ (top panels) and \oione\ (bottom panels) emission lines. The pv diagrams are along a slit through the nucleus of \gal\ in PA = 51\arcdeg. The horizontal white lines correspond the recessional velocity of \gal\ adopted by W13: V$_{\rm recc}$ = 1335 \kms.}
\label{fig:stismuse}
\end{figure}

\section{Software and Analysis Methods}
\label{app:methods}

Section \ref{metandsoft} provides a brief overview of the software and methods used in this work. In this appendix, we provide a more detailed description of these. 

The stellar kinematics was measured using the version of the penalized Pixel Fitting Technique \citep[\ppxf;][]{ppxf1,ppxf2} included in the \gist\ \citep{gist}. The mentioned technique selects a linear combination of stellar templates which best-fit the observed galaxy spectrum and delivers stellar moment maps (total intensity, radial velocity, velocity dispersion, and higher moments).
Before running \ppxf, spaxels were Voronoi binned \citep{vorbin}, setting a minimal SNR in each bin of 300 for the \wfm\ datacube and 80 for the \nfm\ datacube. 
The continuum SNR for the Voronoi binning was calculated within the wavelength range \iaa{5700}{5750}, and spaxels with SNR less than $3$ in the continuum were excluded from the binning.
Bins which included spaxels affected by emission from the jet and its knots were excluded from the analysis. 
We used the MILES stellar template library \citep{miles}, that covers \iaa{3525}{7500} and includes the \mgi\ and \nai\ lines. The \gist\ was run on the wavelength range \iaa{5050}{6000} for the \wfm\ datacube, following \citet{sar18} and \citet{ems14}; and the wavelength range \iaa{5240}{5770} for the \nfm\ datacube, in order to avoid the \mgi\ line.

Following the procedure of \citet{sar18}, we chose to use four moments: the velocity (M1), the velocity dispersion (M2), and the third and fourth coefficients of the Gauss-Hermite polynomials, representing the skewness (M3) and kurtosis (M4) of the LOSVD. We used a 6th (10th) degree additive polynomial for the \nfm\ (\wfm) datacube. We emphasize that \gist\ provides the integrated intensity (M0) per spaxel (not per bin). 

The \gist\  includes a module to model the star formation history and derive the weighted mean of the age and metallicity of the templates used to fit the absorption lines. These are also derived via \ppxf, but in a different run from that described above for the kinematics. Here, for both \wfm\ and \nfm\ datacubes, we requested 4 moments to be extracted and used a 4th-degree multiplicative polynomial. The wavelength ranges were the same as used in the stellar kinematics run. 

The ionized gas kinematics can be also modeled with \gist\ using \textit{GANDALF} \citep{gandalf1,gandalf2,gist}. The program separates the contribution of the emission lines from the stellar absorption lines in each bin. This step is performed using the results of the \ppxf\ run as inputs; for our analysis, we used the results from Voronoi Binning. We chose to fit the wavelength range \iaa{4750}{6800}, since it covers the main emission lines in the MUSE range. 

While binning improves the SNR of the spectrum and thus reliability of the resulting moment maps, spatial resolution is lost in key areas of the maps. For this reason, we also use the emission line moment maps which result from a single Gaussian fit to each spectral line in each spaxel's observed spectrum. 

Since nuclear ionized gas velocities in \gal\ reach almost $\sim$1000 \kms, relatively close line complexes, e.g., \fullhanii, are blended in the nucleus. We thus analyzed the principle ionized gas lines in groups whose wavelengths are close together. For each group, a wavelength range that covers all the lines plus continuum was selected. A straight line was fit to the continuum in order to obtain an 'emission-line only' spectrum. 
The intensity, velocity and dispersion maps of all the lines were obtained from a single Gaussian fit to each line in the group. To facilitate the fitting, two conditions were imposed: (a) all lines in a given group have the same velocity and velocity dispersion; and (b) the amplitude ratios of the following lines were fixed: \oiiione/\oiiitwo\ $= 0.35$; \oitwo/\oione\ $= 0.333$; and \niione/\niitwo\ $= 0.34$. 

\section{PV diagrams of \nfm\ filaments} 
\label{app:pvslits}

Figure \ref{fig:apppvnfm} shows the pv diagrams of the NFM filaments presented in Sect. \ref{nfmfil}, but not shown in Fig. \ref{fig:nfmmoms}. 

\begin{figure}[h]
\centering
\includegraphics[scale=0.5,clip,angle=0]{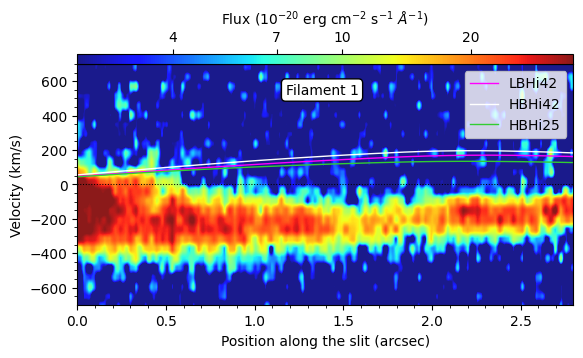} 
\includegraphics[scale=0.5,clip,angle=0]{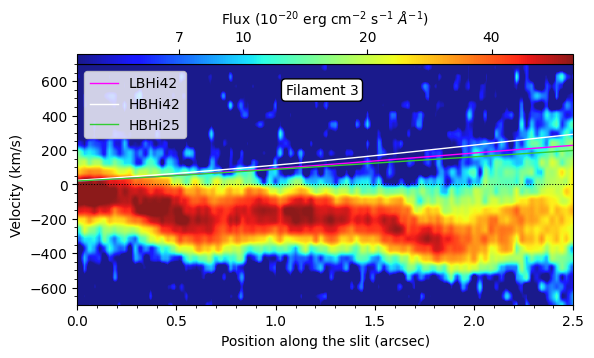} 
\includegraphics[scale=0.5,clip,angle=0]{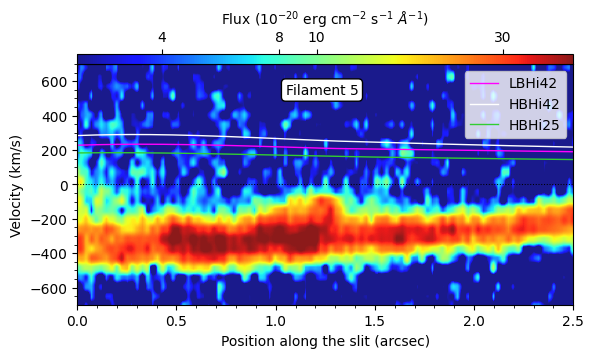} 
\includegraphics[scale=0.5,clip,angle=0]{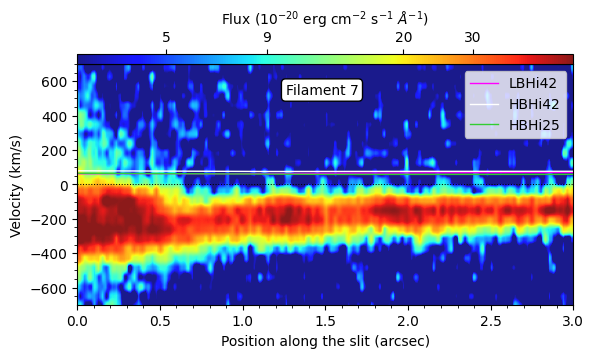} 
\caption{As in Fig. \ref{fig:pvnfm} but for the remaining four of the eight filaments marked in the \nfm\ moment maps (black dotted curves in Fig. \ref{fig:nfmmoms}).
}
\label{fig:apppvnfm}
\end{figure}

\section{Additional masks for the residual velocity maps}

Fig. \ref{fig:maskingapp} shows other masks applied to the residual velocity maps and not shown in Fig. \ref{fig:masking}.

\begin{figure}[h]
\includegraphics[scale=0.3,clip,angle=0]{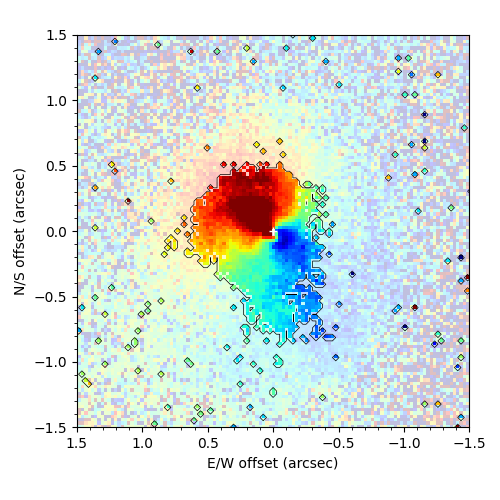}
\includegraphics[scale=0.3,clip,angle=0]{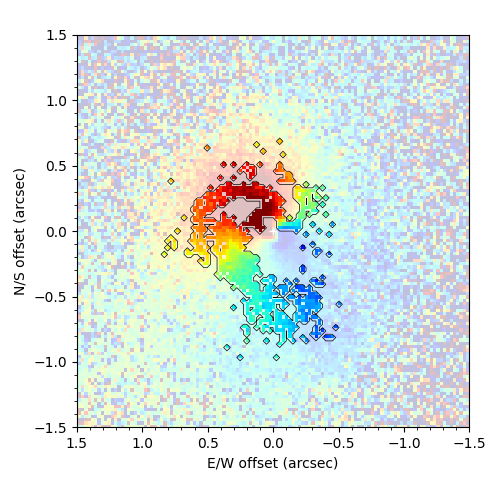}
\includegraphics[scale=0.3,clip,angle=0]{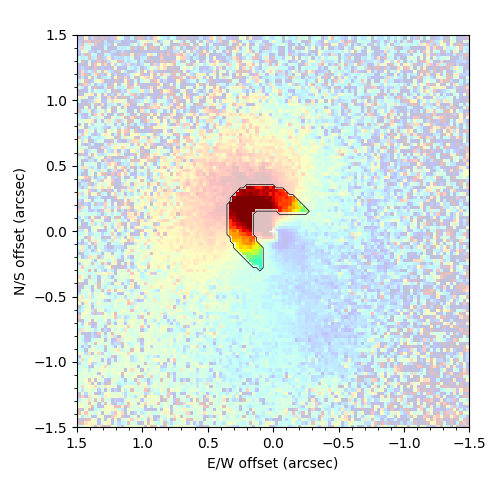}
\includegraphics[scale=0.28,clip,angle=0]{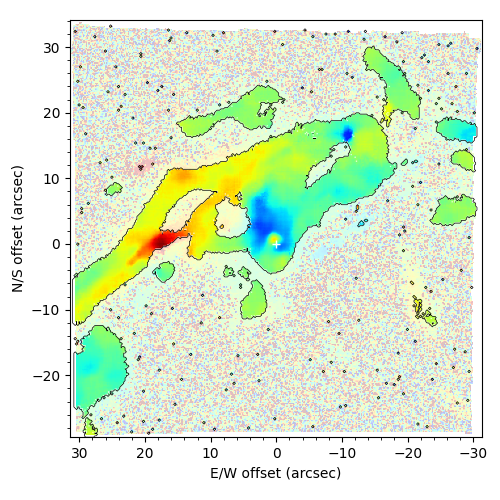}
\centering
\caption{Masks applied to the subtracted velocity maps (see also Fig. \ref{fig:masking}):
\fulloi\ Gauss-fit mask (top left),
\fulloi\ spiral-arm mask (top right),
\fulloi\ annular mask (bottom left),
and \fullhanii\ Gauss-fit mask (bottom right).}
\label{fig:maskingapp}
\end{figure}

\section{Biconical outflow models}

Figure \ref{fig:appbicone} shows two models of a biconical outflow, to illustrate alternative scenarios to the biconical outflow discussed in Sect. \ref{igoutflow} and shown in Fig. \ref{fig:cones}. Here we show the velocity field expected if both cones are fully filled with ionized gas. As the observed blueshifted region to the N is wider than the redshifted region to the S, the half opening angle of the receding cone is illustratively chosen to be smaller than the approaching cone: 
the approaching (receding) cone has a half opening angle of 45$\dg$ (30$\dg$). The model in the left panel creates approximately the observed blue and red shifted areas, but not the velocity gradients within these, and the outflow axis is perpendicular to the jet axis. In the right panel, the cone axes are chosen to project to the PA of the jet, but the resulting velocity field fails to reproduce the observed velocities. 

\begin{figure}[h]
\includegraphics[scale=0.375,clip,angle=0]{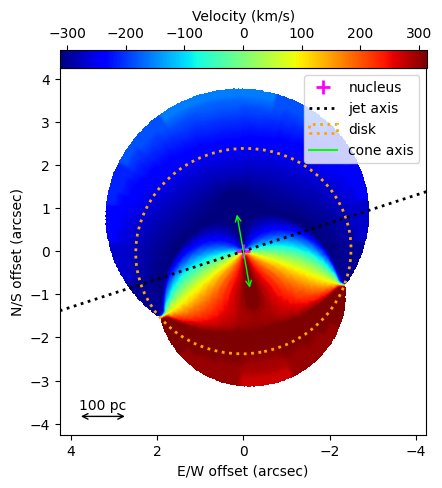}
\includegraphics[scale=0.375,clip,angle=0]{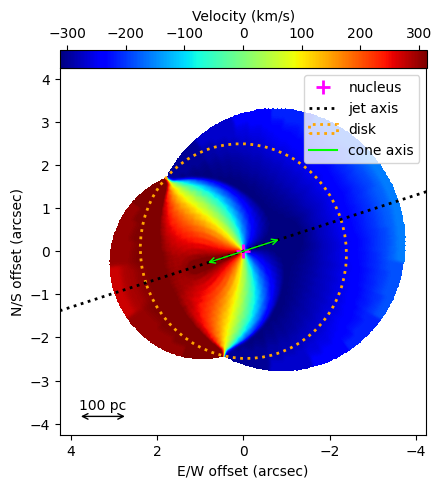}
\centering
\caption{As in the middle and right panels of Fig. \ref{fig:cones}, but this time showing both cones superimposed in a single panel. The left panel illustrates a geometry which allows blueshifts to the N and redshifts to the S, as seen in the left panel of Fig. \ref{fig:cones}. The axes of the cones in this case project to PA 10\arcdeg, roughly perpendicular to the jet axis. The right panel illustrates the case in which the cone axis is aligned with the jet axis, i.e. in PA $-$72\arcdeg.}
\label{fig:appbicone}
\end{figure}

\end{appendix}

\end{document}